\documentclass[aps,prx,reprint,superscriptaddress,longbibliography]{revtex4-2}

\usepackage[T1]{fontenc}
\usepackage{lmodern}
\usepackage{amsmath}
\usepackage{amsthm}
\usepackage{amssymb}
\usepackage{booktabs}
\usepackage{graphicx}
\usepackage{xcolor}
\usepackage[colorlinks=true,allcolors=blue]{hyperref}
\usepackage[capitalise,noabbrev]{cleveref}

\newtheorem{theorem}{Theorem}
\newtheorem{lemma}{Lemma}
\newtheorem{corollary}{Corollary}

\theoremstyle{definition}
\newtheorem{definition}{Definition}

\begin{document}

\title{How Small Can a Single Shor Run Be?\\ Limits and Constructions for secp256k1}

\author{Yicheng Guang}
\affiliation{Department of Computer Science, University of Colorado Boulder, Boulder, Colorado 80309, USA}
\author{Yuliang Xiong}
\affiliation{Department of Computer Science, University of Colorado Boulder, Boulder, Colorado 80309, USA}
\author{Yueqi Chen}
\affiliation{Department of Computer Science, University of Colorado Boulder, Boulder, Colorado 80309, USA}
\author{Xun Gao}
\affiliation{JILA and Department of Physics, University of Colorado Boulder, Boulder, Colorado 80309, USA}

\date{\today}

\begin{abstract}
Most resource estimates for breaking secp256k1 with Shor's algorithm describe one long run per attempt, and the spacetime volume of that run, its peak number of logical qubits times its Toffoli count, sets its logical-error budget.
Eker{\aa}'s tradeoff shrinks each run by using more runs, but how far it can shrink the largest run, and at what circuit cost, has not been determined.
We minimize the volume of the largest run while every key is recovered with probability at least 0.99, within ten times the total volume of three runs of H\"aner et al.\ and $2^{40}$ classical operations.
We prove that a linear run fixed in advance, which starts its group register at a point and discards its label, reveals at most $\log_2(1+(|S|-1)/r)$ bits about a uniform key, where $|S|$ counts its exponent values and $r$ is the group order.
With generic post-processing and the baseline's window costs, splitting alone then shrinks the largest run at most 1.86 times within the budget.
To gain more, we make each exponent bit cheaper with windows that share one table point among exponent values related by six symmetries of the curve, on a point-addition core borrowed from the ECDSA.Fail leaderboard.
One random scalar and one random start per run give every approximate addition uniform operands, and a bound on each run's outcomes certifies, for every key, a decoder that discards flagged runs.
We count every Toffoli gate and every qubit of every run, and each configuration solved all 563 of its pre-registered simulated trials on secp256k1.
The best one shrinks the largest run 3.87 times at 8.79 times the baseline's total volume, against 2.81 times for Eker{\aa}'s tradeoff on the same core, and its run is at least 2.6 times smaller than every design of Initio Intelligence.
\end{abstract}

\maketitle

\section{Introduction}
\label{sec:intro}

Shor's algorithm computes discrete logarithms in polynomial time~\cite{shor1997}, so a large enough fault-tolerant quantum computer could derive an elliptic-curve private key from its public key.
Estimates of how large that computer must be often take secp256k1, the curve that signs Bitcoin transactions, as their case study~\cite{babbush2026,haner2026,long2026ecdsafail}.
A recent estimate for this curve describes one coherent quantum computation of about 1,450 logical qubits and 40 million Toffoli gates~\cite{haner2026}.
We call such a computation, from its first gate until all of its qubits are measured, a \textbf{run}.

A fault-tolerant computer must keep the quantum state of a run intact until the run ends.
Every logical qubit can fail at every step of the run, and one logical error can spoil the result.
In a layout that performs its magic-state operations one after another, the duration of a computation is governed by the number of these operations~\cite{litinski2019game}, and each Toffoli gate consumes a fixed number of magic states.
We therefore measure time in Toffoli gates, as the public ECDSA.Fail leaderboard does for point-addition circuits~\cite{long2026ecdsafail}, and call the product of a run's peak number of logical qubits and its Toffoli count the run's \textbf{volume}.
The error rate per logical qubit and Toffoli gate must then stay well below one over the volume, so that the expected number of logical errors in the run stays well below one.
The volume of a run therefore sets its logical-error budget, and with it the code distance that the run needs.
We judge a computation by $B$, the volume of its largest run, and by $T$, the total volume of all of its runs, retries included.
This paper asks how small $B$ can get while $T$ stays within a fixed budget.

Recent estimates have reduced $B$ by making the arithmetic of a run cheaper~\cite{roetteler2017,haner2020,babbush2026,schrottenloher2026,haner2026}.
In the design of H\"aner et al., 28 windowed point additions dominate the Toffoli count, and each of them processes 16 exponent bits~\cite{haner2026}.
The volume of one point addition keeps falling.
On the ECDSA.Fail leaderboard, for example, open submissions lowered the product of qubits and Toffoli gates of one point addition by 86\%~\cite{long2026ecdsafail}.
Most of these estimates run Shor's algorithm as one full run per attempt, which in the design of H\"aner et al.\ covers 466 exponent bits.
Chevignard et al.\ split the computation into 22 runs on average to save qubits, and each of these runs needs about $2^{39}$ Toffoli gates~\cite{chevignard2026}.
We use the published resource and success estimates of H\"aner et al.\ as our baseline~\cite{haner2026}.
Its reconstructed run volume is $B_0 = 5.72\times 10^{10}$, the product of 1,457 qubits and 39.3 million Toffoli gates.

A natural way to shrink the largest run is to split the computation into more runs, each of which superposes fewer exponent values and learns only part of the key.
Eker{\aa}'s tradeoff algorithm does this, and it evaluates up to a factor of two fewer group operations per run than Shor's algorithm, at the expense of more runs~\cite{ekera2021}.
H\"aner et al.\ note that executing more instances of the algorithm could reduce the size of each instance, but they leave this out for simplicity~\cite{haner2026}.
We determine the limits and the circuit costs of this tradeoff for the largest run on secp256k1 under three budgets.
The total volume stays within ten times the baseline's, every key is found with probability at least 0.99, and the classical post-processing stays within $2^{40}$ operations.

The tradeoff has a limit, because a run that superposes too few exponent values learns almost nothing about the key.
We consider linear runs, which add known points into a group register under the control of exponent qubits, start that register at a point and discard its label.
For such runs we prove that one run reveals at most $\log_2(1+(|S|-1)/r)$ bits about a uniform key, where $S$ is the set of exponent values that the run superposes and $r \approx 2^{256}$ is the order of the group (\cref{sec:limit}).
We combine this bound with generic classical post-processing and the window costs of the baseline.
For runs fixed in advance, the largest run must then keep at least 258.97 exponent bits within the total budget, whatever the sizes of the runs, which takes about 15 windowed additions where the baseline's run takes 28.
Splitting alone therefore shrinks the largest run by at most 1.86 times.
If each run may depend on the outcomes of the earlier ones, the largest run still keeps at least 250 exponent bits within the same budget, and splitting gains at most 1.93 times.
Our construction reduces the largest run further by combining shorter runs with cheaper exponent bits.

Making each exponent bit cheaper while keeping the runs short raises three challenges.
First, a short run reveals only a few bits of the key, so a decoder must combine dozens of runs into the 256-bit key within $2^{40}$ classical operations, even when some of the runs are corrupted (C1).
Second, a windowed addition covers more exponent bits only with a larger table, because a window of $w$ bits selects one of $2^w$ precomputed points and reading such a table costs about $2^w$ Toffoli gates~\cite{gidney2019windowed} (C2).
Third, the cheapest point additions known are approximate and fail on about 0.2\% of their inputs~\cite{long2026ecdsafail}, so the more than a thousand additions of one key recovery must not pull the success probability below 0.99 for any key (C3).

Prior work meets parts of these challenges, but none meets all three.
For C1, Eker{\aa}'s post-processing recovers the key from many short runs by lattice reduction~\cite{ekera2021}, and Ragavan and Vaikuntanathan use lattice reduction to detect and drop the corrupted runs of Regev's factoring algorithm~\cite{ragavan2024}.
For C2, signed windows halve the table~\cite{haner2020}.
The curve secp256k1 has six cheap symmetries, its units, which map a point $(x,y)$ to $(\beta^k x, \pm y)$ for a cube root of unity $\beta$ in the field.
Classical scalar multiplication cuts its precomputation by about five sixths with a digit set that these units map to itself~\cite{xu2024}, but we found no quantum circuit that uses them.
For C3, Proos and Zalka randomize the starting point of the group register to control exceptional additions~\cite{proos2003}.
H\"aner et al.\ also randomize each table with an offset point and bound the failures of approximate arithmetic for every input~\cite{haner2026}.
However, a unit applied to a table point also moves the point's offset.
The offset then depends on the digit, so it no longer shifts every branch of the run by the same known point.

In this work, we build short runs that stay close to the limit and make each exponent bit cheaper.
Our key insight is that, although a random offset point does not commute with the curve's units, a random scalar does.
We multiply every table of a run by one fresh random scalar and start the group register at a uniform random point.
For every exponent input, these choices give each addition a uniform pair of operands along its ideal path, even in windows where one table point stands for six exponent values.

To address C1, each run keeps about 270 exponent bits with smoothly shaped amplitudes, which let the decoder work with a padding of 2 or 3 exponent bits (\cref{sec:shortruns}).
A lattice decoder recovers the key from random subsets of the unflagged runs.
Every decoding attempt pays for its work before it starts, so a key recovery never exceeds $2^{40}$ operations.

For C2, \textbf{unit-orbit windows} store one point for each orbit of the six units, so they cover 18 exponent bits with 43,692 table points, a third of what a signed window of 18 bits needs.
A \textbf{mixed-radix} variant reduces the table to 19,116 points for 16.8 exponent bits.
In a mixed-radix run, the first lookup covers 17 of the 267.88 exponent bits, so the 15 additions must cover 16.73 bits each on average.
Windows of 16.807 bits meet this average with tables less than half the size of those of 18-bit windows.
Every addition runs on a point-addition core that we borrow from the public ECDSA.Fail leaderboard~\cite{odinfree2026,long2026ecdsafail}.
The final addition offers a further saving, because its output can be measured.
Our \textbf{root-sign last addition} measures the $x$-coordinate of the final point and selects its $y$-coordinate by a check on the slope register, with 56\% of the Toffoli gates of a full window.

\begin{figure*}[t]
\centering
\includegraphics[width=\linewidth]{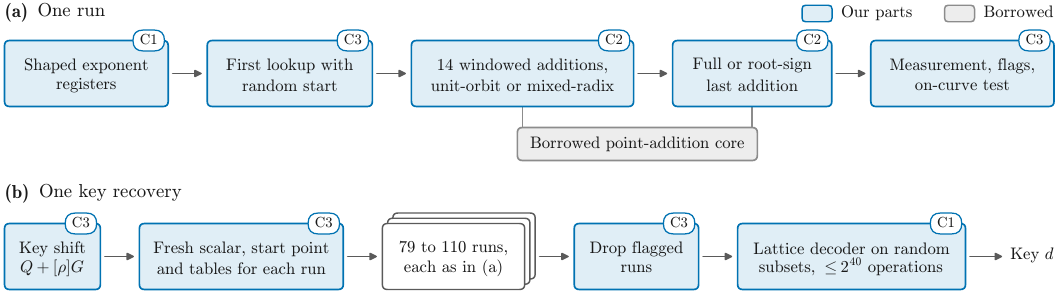}
\caption{Parts of one key recovery and how they connect. The tags C1 to C3 name the challenge that each part meets.}
\label{fig:overview}
\end{figure*}

For C3, the \textbf{run-wide scalar} and the random start make every addition fail with the same probability for every input.
Circuit flags, an on-curve test of the final point and an abort rule identify runs to discard, and the certificate also accounts for errors among the retained runs.
Before each key recovery, we replace the public key $Q$ by $Q+[\rho]G$ for a uniform random $\rho$.
Subtracting $\rho$ from the recovered logarithm gives the same success probability for every key.
Our run lemma bounds the probability of each outcome of a run from the measured failure rates, for window circuits whose error law, averaged over the measurement record, depends only on their operands.
A theorem on the decoder then combines these bounds over the runs into a lower bound on the probability of recovering every key (\cref{sec:certificate}).
\Cref{fig:overview} shows how these parts connect in one run and in one key recovery.

We count every gate of every run and examine the construction in two ways.
A reversible-circuit checker runs every point addition at 256 bits on random and exceptional inputs and counts its Toffoli gates and its peak qubits.
An end-to-end harness then simulates whole key recoveries on secp256k1 and on smaller curves of 16 to 128 bits.
The harness samples the ideal outcome of each run and applies the certificate's outcome-dependent keep rule.
It flags runs at the certified lower bound on visible flags and treats every other run that the certificate does not match to an ideal draw as adversarial.
We fixed the number of runs, the number of trials and the random seed of each configuration before its first trial.
Every addition is charged at a hard cap on the Toffoli gates it executes, and every key recovery stops at $2^{40}$ classical operations, the work of building its tables included.
We compare against the run of H\"aner et al., which succeeds with probability 0.861 per attempt by their heuristic estimate, so three attempts reach 0.997 and the baseline's total volume is $T_0 = 3B_0$.

The best configuration combines mixed-radix windows with the root-sign last addition and uses 102 runs of 1,246 qubits and 11.9 million Toffoli gates.
Its largest run is 3.87 times smaller than the baseline's ($B = 0.2585\,B_0$), at a total volume of $T = 8.79\,T_0$.
With unit-orbit windows and a full last window, 79 runs suffice, and the largest run is 3.64 times smaller ($B = 0.2751\,B_0$) at $T = 7.24\,T_0$.
Each configuration solved all 563 of its trials on secp256k1.
For each configuration, the trials and the Monte Carlo bounds give a lower bound of 0.9945 on the success probability at a confidence of at least 95\%.
An independent audit reproduced the resource totals and the statistical calculations from the saved evidence.

To separate our share of the gain from that of the borrowed core and of the tradeoff itself, we also price Eker{\aa}'s tradeoff on the same core with plain signed windows and plain decoding.
It shrinks the largest run 2.81 times at $T = 7.71\,T_0$, and our windows, decoder and certificate shrink it 1.29 to 1.38 times further.
An earlier audited configuration, whose point addition is entirely our own circuit, reaches 2.14 times at $T = 3.90\,T_0$.

In summary, this paper makes the following contributions.
\begin{itemize}
\item We prove that one linear run of Shor's algorithm reveals at most $\log_2(1+(|S|-1)/r)$ bits about the key, and we derive from it how far splitting alone can shrink the largest run for secp256k1 within a total budget.
To our knowledge, no per-run information bound for Shor-type discrete-logarithm algorithms has been published.
\item We build short runs whose windows use the curve's six units on a borrowed point-addition core, and we give a certificate that bounds their success probability from below for every key although their point additions are approximate.
\item We count every gate of these runs, check whole key recoveries in pre-registered simulated trials on secp256k1, and compare them with unsplit runs and with Eker{\aa}'s tradeoff on the same core.
\end{itemize}

\noindent On QArena~\cite{qarena}, a platform for open challenges in quantum computing, readers can compare designs in the secp256k1 challenge~\cite{qarena-secpsplit} by the volume of their largest run.

\section{Background}
\label{sec:background}

To answer how far splitting can shrink the largest run, we first fix what a run computes, what it costs and how its success is counted.
In this section, we first describe a run of Shor's algorithm and its windowed point additions.
Then, we describe the approximate additions that make a run cheap and the certificates that bound their failures.
Finally, we define the volume and the budgets of a computation, the baseline, and the short runs whose limit \cref{sec:limit} derives.

\subsection{Runs of Shor's algorithm}
\label{sec:bg-runs}

The curve secp256k1 is $y^2 = x^3+7$ over the prime field $\mathbb{F}_p$ with $p = 2^{256}-2^{32}-977$.
Its points form a group of prime order $r \approx 2^{256}$ with a fixed generator $G$, and a public key is the point $Q=[d]G$ of a secret key $d \in \mathbb{Z}_r$.
A run of Shor's algorithm prepares two exponent registers $a$ and $b$ in superposition, adds the point $[a]G-[b]Q$ into a \textbf{group register}, and measures the exponent registers after a quantum Fourier transform (QFT)~\cite{shor1997,proos2003}.
When we describe the circuits that add points into this register, we call it the \textbf{accumulator}.
Each exponent qubit controls the addition of a known point, such as $[2^i]G$ or $-[2^i]Q$, which a classical computer computes before the run.
We call $D = \log_2|S|$ the run's \textbf{exponent bits}, where $S$ is the set of exponent values that the run superposes, so a run of binary registers has as many exponent bits as exponent qubits.
The outcomes of the QFT depend on $d$, and classical post-processing turns the outcomes of one or more runs into $d$.

The QFT followed by a measurement can be carried out one qubit at a time, with classically controlled phase rotations in place of controlled ones~\cite{griffiths1996}.
A run therefore holds only the exponent qubits that the current addition reads, and it measures them before the next addition starts.
As a result, the peak number of qubits of a run is that of the accumulator, the workspace of one addition and the exponent qubits that one addition reads.

\subsection{Windowed point additions}
\label{sec:bg-windows}

Adding $[a]G$ one exponent bit at a time takes one point addition per bit.
Windowed arithmetic adds $w$ exponent bits at once~\cite{gidney2019windowed}.
The $w$ bits address a table of $2^w$ precomputed points, a table lookup loads the selected point into a register, the circuit adds that point into the accumulator, and an unlookup erases the register.
A lookup over $2^w$ entries costs about $2^w$ Toffoli gates, so the window size trades the cost of the lookups against the number of additions.
Signed windows store the points of one sign only and negate the loaded point under the control of one qubit, which halves the table~\cite{haner2020}.
The first window needs no addition, because a direct lookup can load its point into the empty accumulator.

The run of H\"aner et al.\ follows this pattern~\cite{haner2026}.
An initial lookup over $2^{18}$ entries is followed by 28 windowed additions of 16 exponent bits each, so the run covers 466 exponent bits.
Each addition costs about 1.2 million Toffoli gates plus three lookups of $2^{16}$ entries.
The 28 additions therefore take 99.3\% of the run's 39.3 million Toffoli gates, and a run's volume is almost exactly its number of windowed additions times the volume of one addition.

\subsection{Approximate point additions}
\label{sec:bg-approx}

A point addition divides by the difference of two $x$-coordinates, and the division dominates its cost.
The cheapest circuits compute this inverse with walks of a fixed number of steps that suffice for almost every input, so they fail on a small fraction of inputs~\cite{long2026ecdsafail,haner2026}.
The ECDSA.Fail challenge paper, for example, reports that one of its approximate addition circuits fails on 0.192\% of 100,000 independent inputs~\cite{long2026ecdsafail}.
A run executes more than a dozen additions and a key recovery more than a thousand, so a design must bound what their failures do to the success probability, for every input.

Several known techniques give such bounds.
Proos and Zalka start the accumulator at a uniformly random point, so that the exceptional cases of point addition hit a uniformly random fraction of inputs~\cite{proos2003}.
H\"aner et al.\ add a fresh random offset point to every table and a fresh random mask to every approximate adder~\cite{haner2026}.
Every addition then sees a uniformly random pair of operands for every input, and averaging the run's state over the random choices bounds the success probability from below by $\bigl(\max\{0,\sqrt{P_0}-2p_f\}\bigr)^2$, where $P_0$ is the ideal success probability and $p_f$ bounds the failure probability for each input.

\subsection{Volume, success and the baseline}
\label{sec:bg-cost}

Recall that the volume of a run is its peak number of logical qubits times its Toffoli count.
For a computation, $B$ is the largest volume among its runs, and $T$ is the sum over a finite budget of runs fixed in advance, failed runs and retries included.
An approximate addition executes a number of Toffoli gates that depends on its measurement outcomes, so we charge each addition at a hard cap on its executed count.
An addition whose count would pass the cap aborts and flags its run as failed.
The baseline's figure is also an executed count, because H\"aner et al.\ report the largest Toffoli count that they observed in 100 runs of their addition~\cite{haner2026}.

We rebuild the baseline's volume from the figures in the text of H\"aner et al.\ as $B_0 = 1{,}457 \times 39{,}255{,}168 = 5.7195\times 10^{10}$.\footnote{The text gives the Toffoli count of an addition as 1.196 million, rounded, and the runtime but not the Toffoli count of the initial lookup, which we charge at $2^{18}$. The most conservative reading, at the lower rounding edge and without the initial lookup, gives a volume 0.7\% smaller and raises every ratio $B/B_0$ and $T/T_0$ in this paper by 0.7\%.}
Their run succeeds with probability 0.861 per attempt by their heuristic estimate, so three attempts reach $1-0.139^3 = 0.997$ and the baseline's total volume is $T_0 = 3B_0$.
Their rigorous bound of 0.553 per attempt would need six attempts, so $T_0 = 3B_0$ is the stricter reference.

Every computation in this paper must meet three budgets.
First, it must find every key with probability at least 0.99.
We take this target from Eker{\aa}'s estimates of the number of runs, which ask for a success probability of at least 0.99~\cite[Table~2]{ekera2019revisiting}.
A different target can change the number of runs and the total volume, but not the volume of any run of a fixed circuit.
Second, its total volume must stay within $T \le 10\,T_0$.
Third, it may spend at most $2^{40}$ classical operations per key, counting group operations, field multiplications and the word operations of lattice reduction, the precomputation of its tables included.

\subsection{Short runs and the classical budget}
\label{sec:bg-short}

Eker{\aa}'s tradeoff algorithm shortens the exponent registers of a run and makes up for them with more runs~\cite{ekera2021}.
In a group of known order, such as the group of secp256k1, a run with tradeoff factor $s$ superposes $a$ over $256+\sigma$ bits, where $\sigma$ is a small padding, and $b$ over $\ell = \lceil 256/s\rceil$ bits~\cite{ekera2019revisiting}.
It therefore keeps $D = 256+\sigma+\ell$ exponent bits, against 466 in the run of H\"aner et al.
Each run yields an approximate relation $zd \equiv \theta + e \pmod r$, where $z$ and $\theta$ are computed from its outcome and the unknown error $e$ has about $256-\ell$ bits.
The relation therefore carries about $\ell$ bits about $d$, and lattice reduction combines the relations of several runs into $d$.
These approximate relations must be combined within the classical budget, which also limits how far splitting can go.

A generic classical algorithm learns about $d$ only from collisions among the group elements that it computes.
With $W$ group operations it holds $W+2$ elements, counting $G$ and $Q$, and each of their $W(W+1)/2+W+1$ pairs, unless its two elements are the same combination of $G$ and $Q$, collides for at most one value of $d$~\cite{shoup1997,maurer2005}.
With $W = 2^{40}$, the post-processing can therefore single out the key among at most about $2^{80}$ candidates, as a kangaroo search over an interval of $2^{80}$ values does.
Whatever the runs reveal must narrow the key down to a list of about $2^{80}$ candidates.
The question for \cref{sec:limit} is how many exponent bits a run must keep to reveal that much, because that number decides how far splitting can go.

\section{How far splitting can go}
\label{sec:limit}

With runs and their volumes defined, we ask how many exponent bits a run must keep, because that number decides how far splitting can go.
We first state the class of runs that our bound covers.
Then, we bound the information that one run reveals and turn the bound into a floor on the exponent bits of the largest run inside the budget.
Finally, we report how much of the bound the QFT measurement extracts on the distributions we tested, and we name the runs that lie outside the model.

\subsection{Run model}
\label{sec:limit-model}

The bound covers runs that only add known points under the control of exponent qubits.

\begin{definition}[Linear run]
\label{def:linear}
Let $G$ generate a group of prime order $r$, and let $Q=[d]G$.
A \textbf{linear run} fixes, before it starts, a finite set $S \subset \mathbb{Z}^k$ of exponent vectors, complex amplitudes $(\alpha_z)_{z\in S}$ with $\sum_z |\alpha_z|^2 = 1$, and generators $P_i = [u_i + v_i d]G$ for $i=1,\dots,k$ with known $u_i, v_i \in \mathbb{Z}_r$.
It prepares the state $\sum_{z\in S} \alpha_z\, |z\rangle\, |[f_d(z)]G\rangle$ with $f_d(z) = \sum_i z_i (u_i + v_i d) \bmod r$ by controlled additions of known points into a group register that starts at a point of known discrete logarithm.
At the end, the run discards the group register, or measures it and never uses the measured point, and it measures the exponent register by any measurement.
The set, the amplitudes, the numbers $u_i$ and $v_i$ and this measurement depend on $d$ and on its encoding $Q$ only through the outcomes of earlier runs, and they may use randomness that is drawn independently of $d$ and known to the post-processing.
\end{definition}

Shor's run, the runs of Eker{\aa}'s tradeoff and the runs of \cref{sec:construction}, with ideal additions, are linear runs.
Their generators are classical combinations of $G$, $Q$, $[\lambda]G$ and $[\lambda]Q$, where $\lambda$ is the known eigenvalue of the curve's endomorphism, and every such combination has the form $[u+vd]G$.
Apart from its generators and the outcomes of earlier runs, a linear run uses nothing that depends on $d$, so its set cannot, for example, hold only the vector that encodes $Q$, which would reveal all of $d$.
The random start, the run-wide scalar and the key shift of our runs are randomness that \cref{def:linear} allows.
We bound what a computation learns from such runs under four further assumptions.
First, no quantum state passes from one run to the next, so runs fixed in advance, each with its own random choices, are independent given $d$.
Second, the key $d$ is uniform in $\mathbb{Z}_r$.
A key recovery that succeeds with probability 0.99 for every key also succeeds with probability 0.99 for a uniform key, so a bound for a uniform key applies to it.
Third, the post-processing is generic.
It outputs a list of $K$ candidates computed from the outcomes and the randomness only, and $2^{40}$ classical operations give $K \le 2^{80}$ (\cref{sec:bg-short}).
Finally, for the numbers of \cref{sec:limit-floor}, the arithmetic is fixed at the baseline's.
A run pays the baseline's volume for every window of 16 exponent bits after its first 16, which the initial lookup loads at no charge.
Its outcome costs at least one classical operation to read, so a computation within the classical budget makes at most $2^{40}$ runs.

\subsection{One run's information}
\label{sec:limit-thm}

Our bound says that a run must superpose about $r$ exponent values to learn even one bit about the key.
To state it, call two exponent vectors $z$ and $z'$ equivalent when $u\cdot(z-z') \equiv 0$ and $v\cdot(z-z') \equiv 0 \pmod r$, so that they map to the same point for every key, and let $N \le |S|$ be the number of classes of $S$.

\begin{theorem}
\label{thm:info}
For a linear run fixed in advance and a uniform key $d$, the Holevo information between $d$ and the exponent register is at most $\ln(1+(N-1)/r)$ nats.
Every measurement on one run therefore reveals at most $\log_2(1+(|S|-1)/r)$ bits about $d$.
\end{theorem}

The proof compares the state of the run with a reference state that does not depend on $d$.
Two different classes map to the same point for at most one value of $d$, because $f_d(z) = f_d(z')$ is a linear equation in $d$ with at most one root modulo the prime $r$.
The reference state keeps the classes and drops every interference between them.
The Holevo information is at most the average relative entropy between the run's state and this reference, and the relative entropy is at most the R\'enyi divergence of order two, $\ln(1+x_d)$.
Here $x_d = \sum_k \pi_k\,(n_k(d)-1)$ weights each class $k$, of weight $\pi_k$, by the number $n_k(d)-1$ of other classes that share its point at $d$.
Each pair of classes collides for at most one $d$, so $x_d$ averages to at most $(N-1)/r$, and Jensen's inequality finishes the proof (\cref{app:proofs}).

With several runs fixed in advance, the information adds up, and Fano's inequality turns the total into a bound on the success probability.

\begin{corollary}
\label{cor:fano}
Let $R$ linear runs with $N_1,\dots,N_R$ classes be fixed in advance and independent given $d$, and let the post-processing output a list of $K$ candidates.
Then the probability that the list contains a uniform key is at most
\[
P \le \frac{\sum_{t=1}^{R} \log_2\!\left(1+(N_t-1)/r\right) + 1}{\log_2 r - \log_2 K}.
\]
\end{corollary}

A computation may also choose each run after it sees the outcomes of the earlier ones.
The additive form of \cref{cor:fano} fails then, because a later run acts on a narrower posterior and can learn more than a fresh one.
A weaker bound still holds, by a hybrid argument that replaces the runs one by one with the $d$-independent reference (\cref{app:proofs}).

\begin{theorem}
\label{thm:adaptive}
Let a computation choose the set, the amplitudes, the numbers $u_i$ and $v_i$ and the measurement of each of its $R$ linear runs as any function of the earlier outcomes and of randomness drawn independently of $d$, with at most $N$ classes per run, and let it output a list of $K$ candidates.
Then the probability that the list contains a uniform key is at most $K/r + R(N-1)/r$.
\end{theorem}

\subsection{The floor inside the budget}
\label{sec:limit-floor}

\begin{figure*}[t]
\centering
\includegraphics[width=\linewidth]{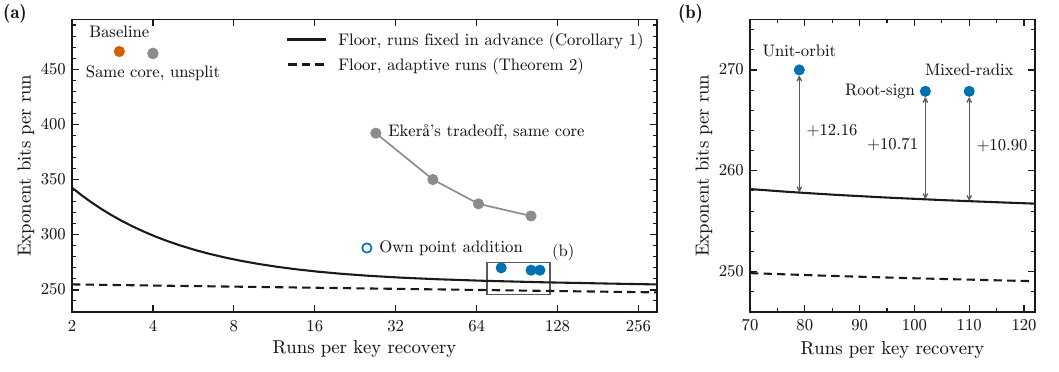}
\caption{Exponent bits per run against the number of runs at $n=256$. (a) The runs that we priced and the floors of \cref{cor:fano} and \cref{thm:adaptive}. (b) The boxed region of (a), with the distance of each of our configurations above the floor of \cref{cor:fano} at its own run count.}
\label{fig:bits}
\end{figure*}

We now put numbers into these bounds for secp256k1.
With success 0.99 and $K = 2^{80}$, \cref{cor:fano} requires $\sum_t \log_2(1+(N_t-1)/r) \ge 0.99\,(256-80)-1 = 173.24$ bits.
For $R$ equal runs, each run must therefore keep at least $256 + \log_2(2^{173.24/R}-1)$ exponent bits.
\cref{tab:floors} lists this floor with its volume at the baseline's arithmetic, and \cref{fig:bits} plots it against the number of runs.

\begin{table*}[t]
\centering
\caption{Floors on the exponent bits of the largest run at $n=256$, for success 0.99 and a list of $2^{80}$ candidates. Volumes are at the baseline's arithmetic, and $T/T_0$ is for $R$ equal runs.}
\label{tab:floors}
\begin{tabular}{rrrrrrr}
\toprule
\textbf{Runs} & \textbf{Bits} & $\boldsymbol{B/B_0}$ & $\boldsymbol{T/T_0}$ & \textbf{Adaptive bits} & \textbf{Adaptive} $\boldsymbol{B/B_0}$ & \textbf{Adaptive} $\boldsymbol{T/T_0}$ \\
\midrule
4 & 299.31 & 0.628 & 0.84 & 253.99 & 0.528 & 0.70 \\
16 & 266.83 & 0.556 & 2.97 & 251.99 & 0.523 & 2.79 \\
21 & 264.24 & 0.550 & 3.85 & 251.59 & 0.522 & 3.66 \\
32 & 261.38 & 0.544 & 5.80 & 250.99 & 0.521 & 5.56 \\
55 & 258.98 & 0.539 & 9.88 & 250.20 & 0.519 & 9.52 \\
64 & 258.47 & 0.538 & 11.47 & 249.99 & 0.519 & 11.07 \\
128 & 256.64 & 0.534 & 22.76 & 248.99 & 0.517 & 22.04 \\
240 & 255.38 & 0.531 & 42.46 & 248.08 & 0.515 & 41.17 \\
\bottomrule
\end{tabular}
\end{table*}

The floor falls slowly with the number of runs, while the total volume grows almost in proportion to it.
At the baseline's arithmetic, a run of 259 exponent bits takes about 15 windowed additions where the baseline's run takes 28, so every run of the floor in \cref{tab:floors} costs more than half of $B_0$.
The budget $T \le 10\,T_0 = 30\,B_0$ therefore caps the number of runs.
At most 55 equal runs fit, and each of them must keep at least 258.97 exponent bits, a volume of $0.5387\,B_0$.

Runs of unequal sizes lower this floor by less than $2^{-78}$ bits.
Beyond its first 16 exponent bits, a run pays the same volume for every further bit, a sixteenth of a window.
The information bound $\log_2(1+(2^D-1)/r)$ instead grows faster with every further bit, because its slope $2^D/(2^D+r-1)$ rises with $D$.
Moving volume from a smaller run to a larger one, up to the size of the largest run, therefore never lowers the total of the information bounds.
For a given largest run, the budget thus buys the largest sum of the information bounds with as many runs of that size as it pays for, one run with the volume left over, and runs of at most 16 bits, which cost no volume.
With 55 runs of 258.98 bits, the run with the leftover volume adds less than $2^{-72}$ bits, and the runs of no volume add less than $2^{-199}$ bits, because a computation makes at most $2^{40}$ runs (\cref{app:frontier}).
Splitting alone therefore shrinks the largest run by at most 1.86 times inside the budget.
Inside a budget of $4\,T_0$, 21 equal runs of 264.24 bits fit, and the limit is 1.82 times.
With no budget on the total volume, the floor falls only slowly, to 255.38 bits at 240 runs, $B \ge 0.530\,B_0$, at a total volume of $42.46\,T_0$.

Adaptive schedules lower the floor by less than a window.
\cref{thm:adaptive} lowers it by 8.48 bits at 64 runs and by 7.30 bits at 240 runs.
The same argument applies to the sum of $(N_t-1)/r$ in its proof when every history stays within the budget, so the largest adaptive run keeps at least 250.15 exponent bits, which 57 equal runs reach, and an adaptive schedule shrinks the largest run by at most 1.93 times.
Our own runs are fixed in advance, so \cref{cor:fano} is the floor that \cref{sec:eval-limit} compares them with.

\subsection{Information that the QFT extracts on tested distributions}
\label{sec:limit-close}

\begin{figure*}[t]
\centering
\includegraphics[width=0.8\linewidth]{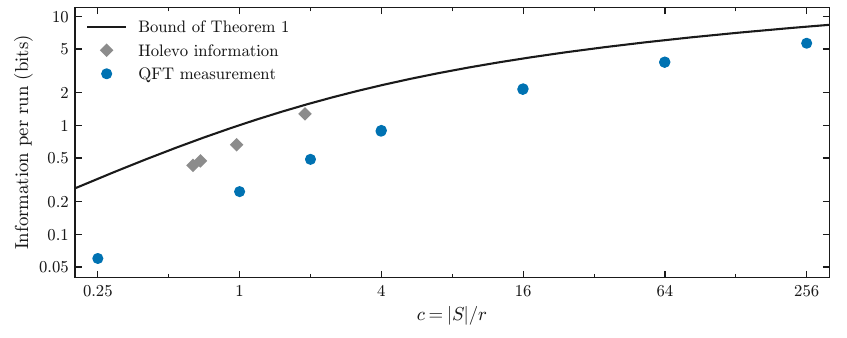}
\caption{One-run information about the key for uniformly weighted two-dimensional boxes of exponent values, with the bound of \cref{thm:info}, against $c = |S|/r$.}
\label{fig:info}
\end{figure*}

\cref{thm:info} bounds every measurement, and the QFT is one particular measurement.
To see how much of the bound the QFT extracts, we computed the exact mutual information between $d$ and the QFT outcome of one run for two-dimensional boxes of exponent values in groups of order 211 to 6841 (\cref{fig:info}).
These exact computations cover the tested boxes, and \cref{sec:eval} evaluates the shaped, mixed-radix runs of \cref{sec:construction} end to end.
On every tested box, the QFT information stays below the bound.
For $c = |S|/r \le 2$ it is about $0.24\,c$ bits, a quarter of the bound at $c=1$, and at $c = 64$ and 256 it lies 2.2 and 2.4 bits below the bound.
The Holevo information of the tested boxes lies between the two, at 61\% to 83\% of the bound for $c$ between 0.63 and 1.9.
Zero-padding the exponent registers before the QFT changed the information by less than 0.005 bits on the boxes we tested, for $c$ between 1 and 64.

\subsection{Runs outside the model}
\label{sec:limit-scope}

Three kinds of runs lie outside \cref{def:linear}.
First, a run may keep a hash of the final point, such as the Legendre symbol of its $x$-coordinate, and uncompute the point, as the runs of Chevignard et al.\ do~\cite{chevignard2026}.
For a hash known to the post-processing, exact computations on small groups give 2.4 to 5.3 times the bound (\cref{app:proofs}), so the requirement that the group register is discarded unread cannot be dropped.
Second, a run may use its measured point in classical group operations or as a generator of a later run.
Third, an algorithm may read the structure of the field or of the curve beyond the known endomorphism.
The bound also covers mid-run measurements whose outcomes control later gates, when deferring them yields a linear run of \cref{def:linear} whose final group register is discarded.
Within the model, splitting alone gains at most 1.86 times inside the budget for runs fixed in advance and 1.93 times for adaptive ones, so a construction that wants more must make each exponent bit cheaper, and it gains most when its runs stay near the floor.

\section{Construction overview}
\label{sec:overview}

\cref{sec:limit} showed that splitting alone gains at most 1.86 times inside the budget, so a smaller run must also make each exponent bit cheaper, and \cref{fig:overview} shows how our construction does both.
Our runs keep about 270 exponent bits with 15 point additions, where the baseline's run keeps 466 with 28, and each addition costs less per exponent bit than the baseline's.

\paragraph{One run.}
In \cref{fig:overview}a, a run first prepares its exponent registers with smoothly shaped amplitudes, so that the decoder can work with short runs (C1).
A first lookup then loads, in one exact table read, the points of the short register and of the top window, shifted by a random start point (C3).
Fourteen windowed additions follow, over unit-orbit or mixed-radix windows whose tables hold one point for each orbit of the curve's six units (C2).
The last addition is either a full window or the root-sign last addition, which measures the final point at a little over half the cost of a full window (C2).
Every addition runs on the point-addition core that we borrow from the ECDSA.Fail leaderboard.
Finally, the run measures its exponent registers and checks its visible flags, which include the core's own flags, an abort cap on the executed Toffoli count and an on-curve test of the measured point (C3).

\paragraph{One key recovery.}
\cref{fig:overview}b shows one key recovery.
Before the runs, the classical side replaces the public key by a random shift of it (C3), and for every run it draws a fresh scalar and a start point and builds the run's tables for the scaled generator (C3).
The 79 to 110 runs then execute one after another or side by side, since they are fixed in advance and no quantum state passes between them.
After the runs, the decoder drops the flagged runs and tries random subsets of the rest with a lattice decoder until one subset yields a key that the public key confirms, all within $2^{40}$ classical operations (C1).

\section{Construction}
\label{sec:construction}

The overview combines short runs, cheaper windows and a certificate for approximate additions.
We now give the cost and the success guarantee of each part, in the order a run meets it.
We first describe the short runs and their first lookup (\cref{sec:shortruns}), then the windows (\cref{sec:windows}) and the borrowed core with the root-sign last addition (\cref{sec:core}).
Finally, we bound the outcomes of each run (\cref{sec:certificate}) and specify the decoder with its success bound and its classical work (\cref{sec:decoder}).

\subsection{Short runs over a two-dimensional exponent}
\label{sec:shortruns}

\paragraph{Two exponent registers.}
Each run follows Eker{\aa}'s tradeoff with a short register $b$ of $\ell$ bits and a register $a$ of $256+\sigma$ bits (\cref{sec:bg-short}).
To let the windows use the curve's units, the register $a$ holds two integers $a_1$ and $a_2$, and the run adds the point $[a_1 + a_2\lambda]G$, where $[\lambda]$ is the curve's endomorphism of \cref{sec:windows}.
A window of $a$ then reads digits of $a_1$ and $a_2$ together.
The unit-orbit configuration keeps $\ell = 11$ and $\sigma = 3$, so each run keeps $D = 270$ exponent bits.
The mixed-radix configurations keep $\ell = 9$ and a register $a$ of 258.88 bits on 261 qubits, so each run superposes $2^{267.88}$ exponent values on 270 qubits.
The decoder estimates $z$ from the two outcomes of $a$, through the point of the lattice $\{(x,y)\in\mathbb{Z}^2 : y \equiv \lambda x \pmod r\}$ nearest to them, and with the outcome of $b$ this gives the run's relation $zd \equiv \theta + e \pmod r$ of \cref{sec:bg-short}.
The nearest point is sometimes the wrong one, and such a run gives a wrong relation, which the decoder of \cref{sec:decoder} must survive.
The exact sampler of \cref{sec:method} draws these runs with their true probability.

\paragraph{Shaped registers.}
With uniform amplitudes, the outcome of a run spreads with heavy tails.
The estimate of $z$ is then wrong with probability about $2^{-\sigma}$ per run, and the errors $e$ have heavy tails, so tradeoffs that decode 20 to 50 runs together need a padding of $\sigma = 10$ to 12 bits.
In our harness on secp256k1, uniform registers solved 37 of 60 trials at $s = 16$ with $\sigma = 10$ and 19 runs, and 30\% of the trials at $s = 32$ with $\sigma = 8$ and 52 runs.
We instead give the register $b$ and each of the two coordinates of the top window of $a$ the amplitudes proportional to $\sin(\pi(x+1/2)/N)$ over the $N$ values $x$ of each register.
They keep the same set of exponent values but make the tails fall much faster, and in the same harness they let the lattice decoder work with $\sigma = 2$ or 3 at $s = 16$ to 20.
The preparations of these states cost 48 Toffoli gates per unit-orbit run and 45 per mixed-radix run.
The top window has 7 bits in the unit-orbit configuration and 8 in the mixed-radix ones.

\paragraph{First lookup.}
The register $b$ and the top window of $a$ are read together by one exact table lookup, which loads their point into the empty accumulator.
The loaded point includes the random start $R_0$ of \cref{sec:certificate}.
This lookup covers 18 exponent bits with 132,380 Toffoli gates in the unit-orbit configuration and 17 bits with 66,588 Toffoli gates in the mixed-radix ones.
Its table is the only one that depends on the public key, and a lookup involves no approximate arithmetic.
As \cref{sec:certificate} shows, this keeps the run's failure constants independent of the key.

\subsection{Unit-orbit and mixed-radix windows}
\label{sec:windows}

\paragraph{The six units of secp256k1.}
The curve secp256k1 has $j$-invariant 0, so the map $\phi(x,y) = (\beta x, y)$, where $\beta$ is a primitive cube root of unity in $\mathbb{F}_p$, sends points to points.
On the group, $\phi$ acts as multiplication by a constant $\lambda$ with $\lambda^2+\lambda+1 \equiv 0 \pmod r$~\cite{glv2001}.
The endomorphisms of the curve form the ring $\mathbb{Z}[\omega]$ with $\omega^2+\omega+1=0$, in which $\omega$ acts as $\phi$.
Its six units $\pm 1$, $\pm\omega$ and $\pm\omega^2$ act as $\pm\phi^k$, so each costs at most one multiplication of $x$ by $\beta$ or $\beta^2$ and at most one negation of $y$ in $\mathbb{F}_p$.
Classical scalar multiplication uses $\phi$ to halve the length of scalars~\cite{glv2001} and to share precomputed points among the six units~\cite{xu2024}.

\paragraph{Unit-orbit windows.}
A unit-orbit window of size $W$ reads $W$ bits of $a_1$ and $W$ bits of $a_2$, which together form a residue $x$ of $\mathbb{Z}[\omega]$ modulo $2^W$.
The window's digit is a chosen element of $\mathbb{Z}[\omega]$ in the class of $x$, and the window adds the precomputed point of that digit.
The prime 2 stays prime in $\mathbb{Z}[\omega]$, so the ring modulo $2^W$ has $4^W$ residues, and the units act on them.
Every nonzero residue lies in an orbit of six, except the three nonzero multiples of $2^{W-1}$, which form one orbit of three.
The table therefore stores one point for each orbit, $(4^W-4)/6+2$ points in all, which is 10,924 at $W=8$ and 43,692 at $W=9$.
Each nonzero residue gets the digit of smallest norm in its class, and the zero residue gets the digit $2^W$ instead of 0, so no addition adds the identity.
An exhaustive check over every residue at $W = 1$ to 7 and $W=9$ confirms the table count and the smallest norm of every nonzero residue's digit.

\begin{figure*}[t]
\centering
\includegraphics[width=0.62\linewidth]{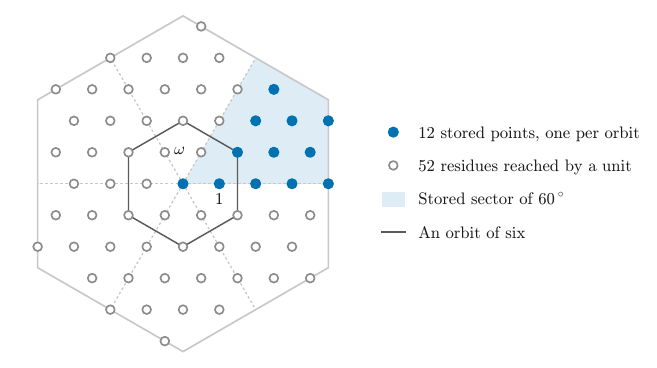}
\caption{The 64 residues of $\mathbb{Z}[\omega]$ modulo $2^3$ and their orbits under the six units.}
\label{fig:orbits}
\end{figure*}

\paragraph{Window circuit.}
The circuit converts the window's residue in place into its orbit representative, a power of $\omega$ and a sign, with 340 Toffoli gates for both directions.
A lookup over the representatives loads $x_T$ and $\beta x_T$ of the representative's point $T$, and the circuit forms the unit's image from them, using $\beta^2 x_T = -x_T - \beta x_T$ and negating $y$ when the sign says so.
The semiclassical QFT needs phase corrections that depend on the integer digit, so the circuit computes the digit into a small register, applies the phases and uncomputes the register, with 486 Toffoli gates at $W=8$ and 532 at $W=9$.
A window of 18 exponent bits thus reads 43,692 table points, a third of the $2^{17}$ that a signed window of 18 bits reads.
In the complex plane the units are the rotations by multiples of 60 degrees, so every orbit of six has one member in each sector of 60 degrees, and the table needs only the residues of one sector (\cref{fig:orbits}).

\paragraph{Mixed-radix windows.}
Unit-orbit windows cover an even number of exponent bits, and a mixed-radix window fills the gap between 16 and 18 bits by adding a seven-valued digit to seven bits of each of $a_1$ and $a_2$.
Its residue is taken modulo $q\,2^7$, where $q = 3+\omega$ or $2-\omega$ is one of the two primes of $\mathbb{Z}[\omega]$ above 7, because 7 splits as $(3+\omega)(2-\omega)$.
The ring modulo $q\,2^7$ is $\mathbb{F}_7 \times \mathbb{Z}[\omega]/(2^7)$, whose second factor is $(\mathbb{Z}/2^7)^2$ as an additive group, so it has $7 \cdot 4^7$ residues, and the window covers 16.807 exponent bits.
The six units act on the nonzero elements of $\mathbb{F}_7$ as all six of them, so a nonzero $\mathbb{F}_7$ part of a residue fixes the unit of its orbit.
The table holds 16,384 orbits whose $\mathbb{F}_7$ part is nonzero and 2,732 whose $\mathbb{F}_7$ part is zero, 19,116 points in all.
The $\mathbb{F}_7$ part sits in a seven-level register on three qubits, which counted circuits prepare and measure after a QFT over $\mathbb{Z}_7$ without any Toffoli gate or ancilla (\cref{app:windows}).
A state-vector check at window sizes 1 to 3 reproduces the outcome law that our sampler uses to $6\times10^{-15}$.

\subsection{The point-addition core and the root-sign last addition}
\label{sec:core}

\paragraph{The borrowed core.}
Every addition runs on a point-addition core that we borrow from the public ECDSA.Fail leaderboard, the leading entry of its main track, submitted by odinfree~\cite{odinfree2026,long2026ecdsafail}.
The leaderboard scores additions of a classical point, and this entry uses 1,227 qubits and 722,917 Toffoli gates on average.
The entry supplies the approximate inversion walks, the multiplication and the squaring of our core.
Its classical addend enters the gate stream only through single-qubit and CNOT gates at five sites, and no Toffoli gate depends on it (\cref{app:windows}).
We therefore turn the core into a windowed addition by loading the window's table point at these five sites.
The loads use select-swap reads~\cite{low2018,berry2019}, which spend a few temporary qubits to read a table with fewer Toffoli gates than a plain lookup, and measurements erase the loaded values.
Our reversible-circuit checker (\cref{sec:method}) runs every kind of window at 256 bits.

\paragraph{Abort caps.}
The core's stream emits 761,147 Toffoli gates, but some of them act only under particular outcomes of the stream's measurements, so the number that one addition executes depends on its measurement record.
We cap the core at 726,000 executed Toffoli gates and charge every window at this cap plus its other Toffoli gates.
An addition that would pass the cap aborts and flags its run.
At these caps a window of $W = 8$ costs 779,515 Toffoli gates on 1,244 qubits, a window of $W=9$ 911,737 on 1,246, and a mixed-radix window 812,589 on 1,245.
In our Monte Carlo of the runs, $5.3\times10^{-6}$ of the additions aborted.

\paragraph{The root-sign last addition.}
The last addition of a run needs no clean output, because the run measures the accumulator right after it.
Our root-sign last addition uses this.
It runs the core's stream up to the square that yields the new $x$-coordinate, with one change that leaves $-x$ of the final point in its register, and measures that register.
The classical side computes the two square roots $\pm Y$ of $x^3+7$ and, for every value of the window's digit, the slopes that the two roots would give with that digit's table point.
A select-swap read over the digit loads 32 bits of the first slope and 16 bits of the second, and the circuit compares them with the low bits of its slope register.
The run accepts only if exactly one of the two slopes matches, which also decides the sign of $y$, and it flags itself otherwise.
Measurements in the $X$ basis then erase the slope register and the loaded bits.
Each of these measurements leaves a known phase that depends on the digit, and the window removes these phases together with the digit-dependent phase corrections that its semiclassical QFT applies anyway.
At $W=9$ this addition costs 513,645 Toffoli gates at its cap on 1,246 qubits, 56\% of a full window of the same size.
Its classical tables take $310 \cdot 4^W + 1{,}032$ operations per run, $2^{26.28}$ at $W=9$, and every key recovery pays for them.
\cref{tab:kinds} adds the digit-register costs, which the counts above leave out, and lists the volume per exponent bit of each kind of addition.

\begin{table*}[t]
\centering
\caption{Kinds of point addition. Toffoli counts are at the abort cap and include the window's digit register. The volume per bit is qubits times Toffoli gates per exponent bit.}
\label{tab:kinds}
\begin{tabular}{lrrrrr}
\toprule
\textbf{Kind} & \textbf{Bits} & \begin{tabular}[b]{@{}r@{}}\textbf{Table}\\ \textbf{points}\end{tabular} & \textbf{Toffoli} & \textbf{Qubits} & \begin{tabular}[b]{@{}r@{}}\textbf{Volume per}\\ \textbf{bit ($10^6$)}\end{tabular} \\
\midrule
Baseline's addition & 16 & 65,536 & 1,392,608 & 1,457 & 126.8 \\
Signed window on the core & 16 & 32,768 & 816,538 & 1,244 & 63.5 \\
Unit-orbit window, $W=8$ & 16 & 10,924 & 780,001 & 1,244 & 60.6 \\
Unit-orbit window, $W=9$ & 18 & 43,692 & 912,269 & 1,246 & 63.1 \\
Mixed-radix window & 16.807 & 19,116 & 813,221 & 1,245 & 60.2 \\
Root-sign last addition, $W=9$ & 18 & 43,692 & 514,177 & 1,246 & 35.6 \\
\bottomrule
\end{tabular}
\end{table*}

\subsection{Equal failure for every input}
\label{sec:certificate}

\paragraph{Why offsets fail on unit-orbit tables.}
Recall that H\"aner et al.\ add a fresh random offset point to every table, so that every addition sees uniform operands~\cite{haner2026}.
A unit-orbit table stores the point $T$ of an orbit's representative, and the circuit adds $u(T)$ for the unit $u$ of the digit's orbit.
If the table stored $T + M$ for an offset point $M$, the circuit would add $u(T) + u(M)$, and the offset $u(M)$ would depend on the digit.
The branches of the run would then differ by offsets that no single known point removes.

\paragraph{The run-wide scalar.}
We instead multiply every table of a run, the first lookup's included, by one scalar $c$ drawn uniformly from $[1, r-1]$, fresh for every run, and we keep a uniformly random start point $R_0$ in the first lookup.
A unit is an element of the commutative ring $\mathbb{Z}[\omega]$, so $u([c]T) = [c]\,u(T)$, and the circuit applies the unit to the scaled point exactly as before.
Along the ideal path of any exponent input $x$, addition $w$ adds the point $B_w = [c]\,T_w(x)$ to the accumulator $A_w = R_0 + [c]\,S_w(x)$, where $T_w(x)$ is the table point that $x$ selects and $S_w(x)$ the ideal partial sum before window $w$.
The accumulator $A_w$ is uniform on the group because $R_0$ is.
The addend $B_w$ is uniform on the group without its identity, because $T_w(x)$ is never the identity, $c$ is uniform and $r$ is prime.
Every addition therefore fails with the same probability for every input, which a Monte Carlo over uniform pairs of operands measures.
The scalar costs no qubit and no Toffoli gate, and the classical side rebuilds the run's tables for $[c]G$ with about $2^{20}$ group operations per run.
The outcome law of the exponent registers does not change, because $c$ only relabels the final value of the group register, so the decoder never needs $c$.

\paragraph{An outcome-wise bound for each run.}
Uniform operands give every input the same failure rate, but a decoder needs more.
It needs each run's outcome to follow a known law except on a small, bounded share of runs.
The bound that provides this needs one property of the circuits.

\begin{definition}[Window interface]
\label{def:interface}
An approximate addition has the \textbf{window interface} if, on every computational basis input, it maps basis states to basis states times a sign, the law of this action over its own uniform measurement record depends only on its operand values, and its good executions are exactly ideal with a sign that depends on the record only.
In particular, the digit enters only as the address of exact reads, and the window's conversion, unit and phase corrections are exact.
\end{definition}

For each addition $w$, let $f_{\mathrm{DO},w}$ be the probability, on uniform operands and a uniform record, that it does not end at the right point with clean ancillas and no flag, and let $f_{\mathrm{P},w}$ be the probability that it ends at the right point, clean and unflagged, but with the opposite phase.
To state the bound for each outcome, we also measure the discarded group register, for the analysis only, in the Fourier basis of the group, and we call its outcome $u \in \mathbb{Z}_r$ the run's \textbf{character label}.
Discarding a register gives the other registers the same law as measuring it in any basis and ignoring the result, so the label changes nothing in the run.
The root-sign last addition measures the end point itself, and for its runs $u$ indexes the terms of a decomposition of each outcome's probability by Parseval's identity (\cref{app:cert}).
In the ideal run, $u$ is uniform, and we write $P_I(o,u) = \kappa(o,u)^2/r$ for the ideal probability of the label $u$ together with the outcome $o$ of the exponent registers, so that $\kappa(o,u)^2$ is the ideal probability of $o$ given $u$.
We also write $x_+ = \max\{x, 0\}$.

\begin{lemma}
\label{lem:outcome}
Let every addition of a run have the window interface, and let the run use the run-wide scalar and the random start.
Then for every key, every outcome $o$ of the exponent registers and every character label $u$,
\begin{gather*}
\Pr[\text{no flag}, o, u] \;\ge\; \frac{1}{r}\left(g_0\,\kappa(o,u) - e_0\right)_+^2,\\
g_0 = 1-\sum_w \left(f_{\mathrm{DO},w} + 2 f_{\mathrm{P},w}\right),
\end{gather*}
where $e_0$, an amplitude-weighted sum defined in \cref{app:cert}, bounds the weight of inputs whose path has two or more failures, recovers after a failure, or ends at a wrong point on the curve.
\end{lemma}

The first-order failure rate enters only through the coherent factor $g_0$, so $1-g_0^2$ is $2\sum_w (f_{\mathrm{DO},w}+2f_{\mathrm{P},w})$ to first order.
Only the remainder $e_0$, which collects paths with two or more failures, recoveries and wrong points on the curve, keeps a factor $1/\kappa$.
The proof averages the run's amplitude over $R_0$, $c$ and the records, as H\"aner et al.\ do for the total success of one run, and it applies Jensen's inequality to each outcome separately (\cref{app:cert}).
Whether each counted window circuit has the window interface is a property of its gates, and \cref{app:a2g} derives it window by window, with every digit layer checked at 256 bits on every residue.
The derivation covers every branch of every input, including the branches that follow a failed addition.
The Monte Carlo that measures these failure rates replays the core's gate stream, and its rerun with the semantics of the counted circuits reproduced every count, so the rates are statistics of these circuits (\cref{app:a2g-replay}).

\paragraph{Visible flags.}
A failed addition is often visible at the end of the run.
The core raises flags on some of its failures, an addition that passes its abort cap stops the run, the root-sign last addition flags a failed slope check, and the measured accumulator fails the curve equation when a failure leaves a value off the curve.
We call a run with any of these signs \textbf{flagged}.

\paragraph{Good, flagged and bad runs.}
\cref{lem:outcome} bounds the probability of every unflagged pair $(o,u)$ from below by the share $K(o,u) = (g_0 - e_0/\kappa(o,u))_+^2$ of its ideal probability $P_I(o,u)$.
Each run can therefore be coupled with an ideal draw of $(o,u)$, which the run reproduces, unflagged, with probability $K(o,u)$ (\cref{app:cert}).
We call a run \textbf{good} when it reproduces its draw, flagged when it shows a flag, and \textbf{bad} otherwise.
A run is good with probability $1-\Delta_0$, where $\Delta_0 = 1-\mathbb{E}_I[K]$, and the outcome of a good run follows the reweighted law $P_I(o,u)\,K(o,u)/(1-\Delta_0)$.
\cref{lem:outcome} says nothing about the outcome of a bad run.
In a separate Monte Carlo of 327,680 paths through 16 windows on the core, no wrong output landed on the curve, and the flagged runs made up about a third of the share $\Delta_0$.
The decoder must therefore survive the remaining unflagged bad runs, which \cref{sec:decoder} turns into a success bound.

\subsection{Decoding the unflagged runs}
\label{sec:decoder}

\paragraph{Decoder protocol.}
The decoder first discards every flagged run.
It then tries distinct, uniformly random subsets of NEED unflagged runs, in an order that depends only on which runs are flagged and on its own coins.
For each subset it runs a lattice decoder, it accepts an answer only if the answer checks against the public key, and it stops at the first answer that checks, or after TRIES attempts or when no untried subset is left.

\begin{theorem}
\label{thm:decoder}
Fix the shifted key, let $R$ runs be independent given it, let each be flagged with probability at least $\varphi_{\mathrm{LB}}$, let $\Delta_0$ be the share of runs that are not good at that key, and let $\nu = \Delta_0 - \varphi_{\mathrm{LB}}$.
Then the decoder fails with probability at most
\[
\sum_{n_F + n_N \le R} \mathrm{Mult}(n_F, n_N; R, \varphi_{\mathrm{LB}}, \nu)\; q(n_N, R - n_F) + f_{\mathrm{good}},
\]
where $q(b, R')$ is the probability that no attempt avoids all $b$ bad runs among $R'$ unflagged runs, and $f_{\mathrm{good}}$ is the probability that one subset of NEED good runs, with outcomes drawn from the reweighted law, fails to decode.
\end{theorem}

The proof rests on one observation.
Turning an unflagged bad run into a flagged one never hurts this decoder, because a flagged run never enters a subset, so the worst case flags runs with probability exactly $\varphi_{\mathrm{LB}}$.
The good runs follow the reweighted law independently of the other runs and of the decoder's order, which sees only the flags (\cref{app:cert}).

\paragraph{The lattice decoder.}
Each run gives a relation $z_i d \equiv \theta_i + e_i \pmod r$ of \cref{sec:bg-short}, with $z_i$ from the register $a$, $\theta_i$ from the register $b$ and an unknown error $e_i$ of about $256-\ell$ bits.
For a subset of NEED runs, the decoder reduces a lattice of dimension $\mathrm{NEED}+1$ built from the $z_i$ and $r$ with LLL~\cite{lll1982} and searches for a lattice vector close to a target built from the $\theta_i$.
The last coordinate of that vector gives the shifted key.
The unit-orbit configuration searches with a branching form of Babai's nearest-plane algorithm~\cite{babai1986} over eight levels and uses subsets of 28 runs.
The mixed-radix configurations follow a failed Babai step with a Schnorr--Euchner enumeration~\cite{schnorr1994} of at most $2^{17}$ nodes and use subsets of 34 runs.

\paragraph{Classical work.}
Every attempt pays for its work before it executes.
LLL is charged at its word operations per swap, and the enumeration pays for each block of 64 nodes before it processes the block's first node.
A key recovery starts at most $\lfloor (2^{40} - w)/\mathrm{cap}\rfloor$ attempts, where $w$ is the work already spent and cap is the charge of one attempt, so it never executes more than $2^{40}$ operations, whatever the outcomes.
The work includes the tables of every run at two group operations per entry and the tables of every root-sign last addition.

\paragraph{The key shift.}
Before the runs, the classical side replaces the public key $Q$ by $Q' = Q+[\rho]G$ for a uniform $\rho$, and it subtracts $\rho$ from the decoded logarithm.
The first lookup is the only table built from $Q'$, and it is exact, so the constants of \cref{lem:outcome} are the same for every key.
For every key $d$, the shifted key $d+\rho$ is uniform, so the failure probability of every key is at most the average of the bound of \cref{thm:decoder} over a uniform shifted key, and the trials of \cref{sec:eval} run the decoder under the same uniform shift.

\paragraph{Constants.}
We measure $f_{\mathrm{DO}}$, $f_{\mathrm{P}}$, $e_0$ and $\varphi_{\mathrm{LB}}$ by a Monte Carlo of whole paths through each configuration's own window sequence, with every addition at its abort cap, and \cref{app:cert} lists them.
Each constant is an upper or lower confidence bound at a level that leaves at least 95\% joint confidence with the trials of \cref{sec:eval}.
With every part counted and bounded, we now describe how we examine the runs.

\section{Methodology}
\label{sec:method}

To examine whether these runs are as small and as reliable as \cref{sec:construction} claims, we count every gate of every run and simulate whole key recoveries against the run of H\"aner et al.
We compare against that run because its qubit count, its Toffoli count and its success probability are all published (\cref{sec:bg-cost}).
To separate the borrowed core's share of the gain from ours, we also price unsplit runs and Eker{\aa}'s tradeoff on the same core.

\paragraph{Circuits and counts.}
A reversible-circuit checker builds each point addition at 256 bits as a stream of $X$, CNOT and Toffoli gates with measurement-based uncomputation.
It runs the stream on random inputs and on the exceptional inputs of point addition (the identity, and points equal or opposite to the addend), and it records every input that ends with a wrong point, a flag, a dirty ancilla or a wrong phase.
It counts the Toffoli gates that the stream emits, the gates it executes under each measurement record, and the peak number of live qubits.
Every window kind, both first lookups and the root-sign last addition passed this check, and the digit registers of the windows were checked as separate circuits and charged in full.
The failure rates of \cref{sec:decoder} need far more inputs than the checker can run, so they come from a bit-sliced classical replay of the core's gate stream that applies the same gate rules to 65,536 inputs at a time.
On the windowed core, the checker found 217 failing inputs out of 98,304, a rate of $2.21\times10^{-3}$, and the replay found 809 out of 327,680, a rate of $2.47\times10^{-3}$.
Following the baseline and the ECDSA.Fail leaderboard, we count Toffoli gates and treat single-qubit rotations as free.
To bound what this leaves out, we allow our smallest run 5,600 single-qubit rotations, above a count of 5,531 that charges 3,238 to its seven-level registers and sine-shaped preparations, 838 to its binary Fourier transforms with every phase as a separate rotation, and 1,455 to the corrections between windows with one rotation for every qubit of the digit registers.
At a precision of $10^{-9}$, these rotations add about 1.5\% to the run's $T$ count (\cref{sec:discussion}).
Physical-level estimates, such as code distances, physical qubit counts and run times, are outside the scope of this paper.

\paragraph{End-to-end harness.}
The harness simulates whole key recoveries, from the key shift to the decoder's answer.
It samples the outcome of each run exactly from the run's ideal outcome law, with work linear in the run's exponent bits.
Discarding the group register gives the exponent registers the same law as measuring it in the Fourier basis of the group, where every addition of a known point acts as a phase on the exponent qubits.
The sampler therefore draws the character label $u$ uniformly and then the outcome of each register given $u$.
For a register of uniform qubits, each outcome bit follows from the earlier ones with a probability $\cos^2$ of an exactly reduced rational.
For a shaped register, the sampler computes the outcome law of the register's amplitudes exactly, and for a seven-level register it uses the law of its Fourier measurement over $\mathbb{Z}_7$.
A test against full state-vector simulation agrees to $2\times10^{-17}$, and the sampler reproduces Eker{\aa}'s exact value of one outcome probability~\cite{ekera2019revisiting} to all 17 printed digits.
The harness then applies the keep rule of the certificate.
It flags each run with probability $\varphi_{\mathrm{LB}}$, keeps the ideal outcome $o$ of an unflagged run, drawn with its label $u$, with probability $K(o,u)/(1-\varphi_{\mathrm{LB}})$ for the share $K$ of \cref{sec:certificate}, which is below 1 because $K \le g_0^2 < 1-\varphi_{\mathrm{LB}}$, and otherwise marks the run bad, so that every decoding attempt on a subset that holds it fails.
The decoder sees points only as opaque labels, never their coordinates or discrete logarithms, and the harness stops any key recovery whose classical work passes $2^{40}$ and counts it as a failure.
Besides secp256k1, the harness runs the same configurations on curves of 16, 24, 32, 64 and 128 bits, with the run counts scaled to each size.
At these smaller sizes the additions are stand-ins with the failure constants of ordinary windows, so the smaller curves check the decoder and the run budget but not the root-sign last addition itself.

\paragraph{Pre-registration and statistics.}
For each configuration, we committed the number of runs, the number of trials and the random seed before its first trial, and the trials ran from a copy of the code frozen at that commit.
Each configuration ran 563 trials on secp256k1 and 200 trials at each smaller size.
With no failure in 563 trials, the exact one-sided bound gives a success probability of at least $0.045^{1/563} > 0.994506$ at confidence 0.955~\cite{clopper1934}, and the Wilson interval~\cite{wilson1927} is $[0.9932, 1]$.
The Monte Carlo bounds of the constants spend another 0.005 of confidence, so for each configuration the joint confidence is at least 95\%.
The seven-level circuits of the mixed-radix register were written and checked after the trials of the mixed-radix configuration with a full last window were frozen.
They use no Toffoli gate and no ancilla, so that configuration's counts are unchanged.

\paragraph{Audit.}
An independent audit examined our three configurations from the saved evidence.
It checked that the run count, the trial count, the seed and the pass rule were committed before the first trial, reconciled the resource totals with the counted components, and reproduced the statistical bounds.
It also read the decoder and the harness code, including the key shift, the Monte Carlo constants of each window sequence and the charging of classical work before it executes, and it ran small checks of its own.
It did not rerun the trials, the circuit checks or the replay of the core's gate stream.
The audit accepted all three configurations.

\paragraph{Unsplit runs on the same core.}
To isolate the core, we price one run in the baseline's own shape on the borrowed core.
It keeps 256 bits of $a$ and 208 bits of $b$, and its post-processing finds the 48 bits of the key that the shorter register $b$ leaves open by an interval search of at most $2^{36.5}$ operations per run.
With plain signed windows of 16 bits, the size that minimizes its volume among 12 to 20 bits, and a 16-bit first lookup, it has 28 additions.
A second unsplit run uses our windows, last additions and shaped first lookup in two registers over $\mathbb{Z}[\omega]$, with 27 additions.
Both runs read approximate additions from tables built from the public key, which \cref{lem:outcome} does not cover.
Their run counts therefore come from the heuristic model of the baseline itself, in which each approximate addition spoils its run with four times its failure rate, and under that model they need four runs each.
The run with signed windows was sampled in the harness under this model, and the run with our windows only in its decoder.

\paragraph{Eker{\aa}'s tradeoff on the same core.}
A third control runs Eker{\aa}'s tradeoff on the same core with plain signed windows and plain decoding~\cite{ekera2019revisiting}.
For a tradeoff factor $s$, a batch holds Eker{\aa}'s number of runs for a 256-bit group, each with a register $a$ of $256+\sigma$ bits and a register $b$ of $\lceil 256/s\rceil$ bits, and LLL with Babai's algorithm decodes the batch as a whole.
A batch that fails is rerun whole, with no subsets, flags or certificate.
Under the same heuristic model, one bad run spoils its batch, so a larger $s$ shortens the runs but adds runs that must all be good.
Among the factors whose total volume stays within $10\,T_0$, $s = 4$ gives the smallest largest run, with batches of 5 runs of 328 bits.
A batch then succeeded in 0.3235 of 2,000 trials, and 13 batches, 65 runs in all, solved 0.9950 of 5,000 trials at a total volume of $7.71\,T_0$.

The code and data are available at \url{https://github.com/a85tract/secp256k1-shor-runs}.

\section{Evaluation}
\label{sec:eval}

Using these counts and trials, we answer four questions about the construction.
\begin{enumerate}
\item How small is the largest run, and at what total volume (\cref{sec:eval-size})?
\item Where does the gain come from (\cref{sec:eval-gain})?
\item How close are the runs to the limit of \cref{sec:limit} (\cref{sec:eval-limit})?
\item What does the success certificate cost (\cref{sec:eval-cert})?
\end{enumerate}

\subsection{How small the largest run gets}
\label{sec:eval-size}

\paragraph{Run volume.}
Our best configuration shrinks the largest run 3.87 times, to $B = 0.2585\,B_0$, at a total volume of $T = 8.79\,T_0$.
It combines mixed-radix windows with the root-sign last addition, and each of its 102 runs has 15 additions, 11,866,224 Toffoli gates and 1,246 qubits (\cref{tab:runs}).
The same windows with a full last window use 110 runs and shrink the largest run 3.74 times, to $0.2672\,B_0$, at $9.80\,T_0$.
Unit-orbit windows with a full last window use only 79 runs and shrink the largest run 3.64 times, to $0.2751\,B_0$, at $7.24\,T_0$, the lowest total of the three.
The run of our best configuration is also 2.71 and 2.76 times smaller than the two runs of Initio Intelligence at the baseline's 1,457 qubits, and 2.64 times smaller than their smallest run, at 1,380 qubits (\cref{tab:published})~\cite{initio2026}.

\paragraph{Recovery success.}
Each configuration solved all 563 of its trials on secp256k1, which bounds its success probability from below by 0.9945 at a joint confidence of at least 95\%.
The run counts of 79, 110 and 102 were registered before the trials (\cref{sec:method}), so the trials show that each count suffices for the target of 0.99, not that it is the smallest count that does.
At the smaller sizes, each configuration solved all 200 trials per size, except one trial at 128 bits for the root-sign configuration (\cref{app:trials} lists the trials by size).
That trial drew 9 bad or flagged runs out of 28, found no clean subset and stopped at its $2^{40}$ budget, which the registered rule for the smaller sizes allows.

\begin{table*}[t]
\centering
\caption{Runs that we priced at $n=256$. Bits is the base-2 logarithm of the number of exponent values that a run superposes, and Additions excludes the first lookup. Success gives the trials solved on secp256k1, or the heuristic model that sets the run count.}
\label{tab:runs}
\resizebox{\linewidth}{!}{%
\begin{tabular}{lrrrrrrrr}
\toprule
\textbf{Run} & \textbf{Runs} & \textbf{Bits} & \textbf{Additions} & \textbf{Qubits} & \textbf{Toffoli} & $\boldsymbol{B/B_0}$ & $\boldsymbol{T/T_0}$ & \textbf{Success} \\
\midrule
Baseline~\cite{haner2026} & 3 & 466 & 28 & 1,457 & 39,255,168 & 1.0000 & 1.00 & heuristic \\
Signed windows, unsplit & 4 & 464 & 28 & 1,244 & 22,896,081 & 0.4980 & 0.66 & model \\
Unit-orbit windows, unsplit & 4 & 464 & 27 & 1,246 & 22,151,891 & 0.4826 & 0.64 & model \\
Mixed-radix windows, unsplit & 4 & 464.5 & 27 & 1,246 & 21,823,943 & 0.4754 & 0.63 & model \\
Mixed-radix and root-sign, unsplit & 4 & 464.5 & 27 & 1,246 & 21,425,851 & 0.4668 & 0.62 & model \\
Eker{\aa}'s tradeoff on the core & 65 & 328 & 20 & 1,244 & 16,363,777 & 0.3559 & 7.71 & model \\
Own point addition, signed windows & 25 & 288 & 17 & 1,340 & 19,954,242 & 0.4675 & 3.90 & 1,000 of 1,000 \\
Unit-orbit & 79 & 270 & 15 & 1,246 & 12,626,051 & 0.2751 & 7.24 & 563 of 563 \\
Mixed-radix & 110 & 267.88 & 15 & 1,246 & 12,264,316 & 0.2672 & 9.80 & 563 of 563 \\
Mixed-radix and root-sign & 102 & 267.88 & 15 & 1,246 & 11,866,224 & 0.2585 & 8.79 & 563 of 563 \\
\bottomrule
\end{tabular}}
\end{table*}

\begin{figure*}[t]
\centering
\includegraphics[width=\linewidth]{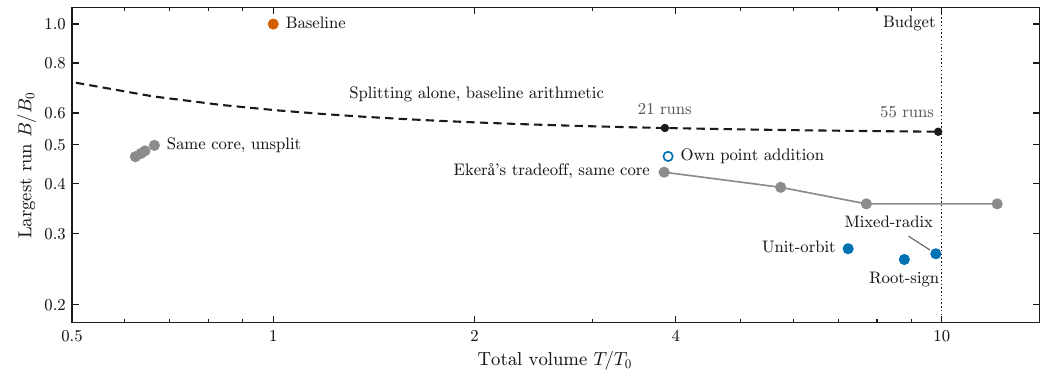}
\caption{Volume of the largest run against the total volume, both relative to the baseline. The dashed curve is the floor of splitting alone at the baseline's arithmetic, and its two marks are the smallest largest runs inside $4\,T_0$ and $10\,T_0$, with 21 and 55 runs.}
\label{fig:frontier}
\end{figure*}

\paragraph{Comparison with the other runs.}
\cref{fig:frontier} places these configurations among the other runs that we priced.
Splitting alone, at the baseline's arithmetic, cannot bring the largest run below $0.5387\,B_0$ inside the budget (\cref{sec:limit-floor}).
The unsplit runs on the borrowed core sit at 0.47 to $0.50\,B_0$ with a total volume below the baseline's, because their four runs together cost less than the baseline's three attempts.
Eker{\aa}'s tradeoff on the same core, with plain signed windows and plain decoding, reaches $0.3559\,B_0$ at $7.71\,T_0$, a largest run 2.81 times smaller than the baseline's.
Our configurations shrink the largest run 1.29 to 1.38 times further within the same budget.
An earlier configuration with our own point addition and signed windows reaches $0.4675\,B_0$ at $3.90\,T_0$, with 1,000 of 1,000 trials and an independent audit, and every gate of its arithmetic is ours.
These results show that the split runs reach a largest run near a quarter of the baseline's inside the budget, at a total volume 11 to 15 times that of the unsplit run with the same windows.

\subsection{Where the gain comes from}
\label{sec:eval-gain}

Of the 3.87-fold gain of our best configuration, the move to the borrowed core, with our select-swap reads, supplies 2.01 times, our windows 1.07 times and splitting 1.81 times.
These factors compare successive stages of one run (\cref{tab:runs}).
The first stage moves the baseline's run onto the borrowed core, with its read layer and a 16-bit first lookup, the second adds our windows, and the third splits the run.

\paragraph{Core.}
In the baseline's own unsplit run, the borrowed core in our signed-window port shrinks the largest run 2.01 times, from $B_0$ to $0.4980\,B_0$.
Against the same port with plain reads, which still has the core's stream and the 16-bit first lookup, the select-swap reads with which we load the table points at the core's five sites add 1.095 times.
The rest of the factor, 1.83 times, comes from the core's gate stream, with a first lookup of 16 bits instead of 18 that accounts for less than 1\% of it, and it belongs to the entry that we borrow.

\paragraph{Windows.}
Our windows add 1.03 to 1.07 times in the same unsplit run.
Unit-orbit windows with a full last window and a shaped first lookup give 1.032 times, mixed-radix windows 1.015 times more, and the root-sign last addition 1.019 times more.
This share is small because the table reads, the only part of a window that the units shrink, take about a tenth of a window's Toffoli gates on the core, 90,535 of the 851,685 Toffoli gates that a signed window of 16 bits emits.
At 16 bits, the six units cut the table from 32,768 points to 10,924 and make its reads cheaper, but each window still performs one point addition on the core.
With the cost of recovering the selected point from its representative included, a unit-orbit window costs 780,001 Toffoli gates at its abort cap against 816,538 for a signed window (\cref{tab:kinds}), 4.5\% less.

\paragraph{Splitting.}
Splitting into 79 to 110 runs adds 1.75 to 1.81 times.
It cuts the additions per run from 27 to 15 and the exponent bits per run from 464 to 270, or from 464.5 to 267.88, a step of 1.72 to 1.73 times in the number of exponent bits.
A window of the borrowed core costs about half of the baseline's, so the budget of $10\,T_0$ pays for about twice as many windows.
On this core, the argument of \cref{sec:limit-floor} allows splitting alone at most 1.87 times, with 112 runs of 256.94 bits.
Our split stays 3.5 to 6.3\% below this limit, with runs 11 to 13 exponent bits above its floor.
Eker{\aa}'s tradeoff on the same core gains 1.40 times over the unsplit run with signed windows, and our windows, decoder and certificate gain 1.29 to 1.38 times over that control.

\paragraph{Physical split.}
In physical terms, the 3.87-fold gain combines 1.73 times fewer exponent qubits per run (466 to 270), 1.92 times fewer Toffoli gates per exponent qubit (84,239 to 43,949) and 1.17 times fewer peak qubits (1,457 to 1,246).
Overall, splitting and the borrowed core each supply about a factor of two, because splitting almost halves the additions per run and the core almost halves the volume of each addition, while our windows make each addition a few percent cheaper.

\subsection{How close the runs are to the limit}
\label{sec:eval-limit}

\paragraph{Gap above the floor.}
At their run counts, our runs keep 10.7 to 12.2 more exponent bits than \cref{cor:fano} requires (\cref{fig:bits}b).
At 79 runs the floor is 257.84 bits, and the unit-orbit configuration keeps 270, 12.16 more.
At 102 runs the floor is 257.17 bits, and the root-sign configuration keeps 267.88, 10.71 more.
At 110 runs the floor is 256.98 bits, 10.90 below the mixed-radix configuration with a full last window.

\paragraph{Where the gap comes from.}
Our decoder differs from the accounting of \cref{cor:fano} in three ways.
First, it decodes subsets of 28 or 34 runs, because one bad run spoils every subset that holds it, whereas \cref{cor:fano} pools the information of all runs of a key recovery.
Second, it recovers all 256 bits of the key from one subset, whereas \cref{cor:fano} leaves a list of $2^{80}$ candidates to classical search and requires only 173.24 bits of information.
Together, these two differences raise the information that each run must carry from 2.2 bits, 173.24 bits over 79 runs, to 9.1 bits, 256 bits over 28 runs, in the unit-orbit configuration.
Third, a run delivers less to the decoder than it keeps.
The padding of the register $a$ fixes the decoder's estimate of the relation but carries no information about $d$, and the outcome error of the shaped register $b$ costs about one bit per run.
Pooling every unflagged run would remove the subsets, but the decoder would then have to tolerate the bad runs inside its input within $2^{40}$ operations, and \cref{thm:decoder} certifies the subset decoder evaluated here.

\subsection{What the success certificate costs}
\label{sec:eval-cert}

\paragraph{Run overhead.}
The run-wide scalar changes only the classical tables, so it adds no qubit and no Toffoli gate, and the flags come from the counted circuits and from classical checks of the measurement record and of the measured accumulator.
The certificate's cost therefore shows in the runs and the classical work of a key recovery.
In the trials, about one run in ten is bad or flagged at 256 bits.
The unit-orbit configuration lost 4,405 of 44,477 runs (9.9\%), the mixed-radix configuration with a full last window 6,046 of 61,930 (9.8\%) and the root-sign configuration 5,302 of 57,426 (9.2\%).
Because one bad run spoils every subset that holds it, a key recovery must keep enough clean runs for a clean subset to come up within the attempts that its classical budget pays for.
Our configurations use 79 runs for subsets of 28 and 102 or 110 runs for subsets of 34, 2.8 to 3.2 times the subset size.

\paragraph{Classical work.}
The classical work stayed below $2^{40}$ in every trial.
The mean key recovery used $2^{32.7}$, $2^{33.3}$ and $2^{34.0}$ operations in the three configurations, and the largest one $2^{39.06}$, $2^{39.85}$ and $2^{37.91}$.
The largest, a mixed-radix trial with 20 bad or flagged runs out of 110, started at least 6,740 of its at most 7,377 attempts before a clean subset decoded, because every subset that holds a bad run costs a full attempt.
No single attempt reached its cap.
With the root-sign last addition, every key recovery pays $2^{32.95}$ for the tables of its last additions, so its smallest trial used $2^{32.98}$.

\paragraph{Run counts of the three configurations.}
Mixed-radix windows cut the largest run by 2.9\% against unit-orbit windows, and the root-sign last addition cuts it by 3.2\% more.
Both mixed-radix configurations use more runs than the unit-orbit configuration, 102 or 110 against 79.
We attribute this to their shorter register $b$, 9 bits against 11, which carries fewer bits per run and needs subsets of 34 runs instead of 28.
The root-sign configuration uses fewer runs than its sibling with a full last window, 102 against 110, because its last addition fails less often, $1.6\times10^{-3}$ against $2.2\times10^{-3}$, and leaves fewer bad or flagged runs, 9.2\% against 9.8\%.
Overall, the certificate keeps the success probability of a whole key recovery at 0.99 or more for every key without adding a qubit or a Toffoli gate to any run, and the bad runs that it tolerates cost about three runs for every run of the decoded subset and classical work below $2^{40}$.

\section{Discussion}
\label{sec:discussion}

The measured gains combine shorter runs with cheaper additions.
We now examine which parts of the success bound rest on premises beyond the counts, which savings the tested designs leave, and which parts of the construction carry over to other curves.

\paragraph{Scope of the evidence.}
Our numbers are logical-level counts and exact sampling of each run's ideal outcome law, and the success bound adds two premises to them.
The first premise is the window interface of \cref{def:interface}.
The reversible-circuit checker examines every window on random and exceptional inputs, but the interface is a statement about all inputs, so \cref{app:a2g} derives it from each window's gates, on every branch of every input, and checks every digit layer at 256 bits on every residue.
The second premise is that the constants of \cref{lem:outcome} hold at their stated confidence.
They come from the classical replay of the core's gate stream, whose failure rate on the windowed core, $2.47\times10^{-3}$ over 327,680 inputs, was the higher of the two when we compared it with the checker's $2.21\times10^{-3}$ over 98,304 inputs (\cref{sec:method}).
The Monte Carlo of whole paths then bounds $e_0$ and $\varphi_{\mathrm{LB}}$, so the success bound is a bound from sampling.
We reran this Monte Carlo with the semantics of the counted circuits, and all nine batches, 1,179,648 paths, gave the same counts, so the constants are statistics of the circuits as built (\cref{app:a2g-replay}).
Two further items lie outside the counts.
Each mixed-radix window uses 291 single-qubit rotations, 7 to prepare its seven-level register and 284 to measure it, which our Toffoli convention treats as free.
A fault-tolerant build must synthesize them, and an error of $\epsilon$ per rotation changes the operation of each window by at most $291\,\epsilon$ in operator norm.
At $\epsilon = 10^{-9}$, Selinger's method synthesizes a rotation with about $10+4\log_2(1/\epsilon) \approx 130$ $T$ gates~\cite{selinger2015}, and a Toffoli gate takes four $T$ gates with measurement and feedforward~\cite{jones2013}.
The 5,600 rotations that we allow our smallest run (\cref{sec:method}) then add about 728,000 $T$ gates to its 47.5 million, or 1.5\%.
They also change the run's operation by at most $5.6\times10^{-6}$ in operator norm, so the outcomes of a key recovery of 102 runs move by at most $1.2\times10^{-3}$ in total variation, inside the margin of 0.0045 between the success bound and 0.99.
The second item is that the run counts of the unsplit runs and of Eker{\aa}'s tradeoff on the same core come from a heuristic model, as the baseline's own success estimate does.

\paragraph{Remaining savings in the tested designs.}
Among the further levers that we priced outside the core of our best configuration, the size of the last addition, the mix of windows, the norm of the mixed-radix windows, the padding, the decoder and an added flag were each worth 0.3\% of $B$ or less.
The core's own gate stream, in contrast, accounts for 761,147 of the 851,685 Toffoli gates that a signed window on the core emits.
A newer entry on the leaderboard would shrink the core's share of every addition, most of its Toffoli gates, in proportion to its saving, and the best entry on 2026-11-13 was 0.03\% below the one we use.
Cheaper point additions are therefore the largest saving that our designs leave.

\paragraph{Directions that do not help.}
\cref{app:closed} lists the other directions that we examined, with their results and the grade of their evidence.
Other run shapes, several instances per run and reductions to related problems still add points of the form $[u+vd]G$ into one discarded group register, so \cref{thm:info} applies to them, and \cref{thm:adaptive} to adaptive chains of them.
On the small groups that we computed exactly, zero-padding the exponent registers changed the information by less than 0.005 bits, and at $c = 64$ and 256 the QFT came within 2.2 and 2.4 bits of the bound that every measurement obeys (\cref{sec:limit-close}).
Inversion circuits with compressed or recomputed transcripts traded qubits for Toffoli gates at a flat or worse product, and a four times larger failure budget for the core's walks bought 1.5 to 2\% of $B$.
Among the shorter configurations that we priced, none fits both budgets.
Runs of 266 exponent bits needed 4 to 8 more runs than $10\,T_0$ allows, and runs of 264 bits could not be decoded within $2^{40}$ operations.

\paragraph{Other curves.}
\cref{thm:info} holds in every group of prime order, so the limit of \cref{sec:limit} applies to P-256 and to every other curve of prime order, with that curve's own window costs.
The unit-orbit and mixed-radix windows need the six units of a curve with $j$-invariant 0, so a curve such as P-256 keeps signed windows, and the run-wide scalar commutes with their unit $-1$ as well.
The flags, the outcome-wise lemma and the subset decoder carry over to any curve of prime order unchanged.

\section{Related work}
\label{sec:related}

Our results bound how far splitting can shrink the largest run and build runs that come close to that bound.
The works closest to ours estimate the cost of Shor's algorithm on elliptic curves, split it into several runs, bound what generic quantum algorithms can do, use the curve's endomorphism, or certify approximate arithmetic.

\paragraph{Resource estimates for elliptic-curve discrete logarithms.}
Proos and Zalka gave the first detailed circuits for Shor's algorithm on elliptic curves~\cite{proos2003}.
Roetteler et al.\ later counted the qubits and Toffoli gates of a complete run~\cite{roetteler2017}, H\"aner et al.\ added windows and cheaper arithmetic~\cite{haner2020}, and Litinski cut the Toffoli count further~\cite{litinski2023}.
Recent estimates for 256-bit curves include those of Babbush et al.~\cite{babbush2026}, Schrottenloher~\cite{schrottenloher2026}, Luo et al.~\cite{luo2026} and H\"aner et al.~\cite{haner2026}, and the ECDSA.Fail leaderboard collects open point-addition circuits~\cite{long2026ecdsafail}.
Chevignard, Fouque and Schrottenloher compress the final point of each run to one Legendre-symbol bit, which minimizes the number of qubits, and they split the computation into 22 runs on average~\cite{chevignard2026}.
Initio Intelligence gives circuits for one run in thirteen designs at 1,380 to 2,022 qubits, with 24.9 to 28.7 million Toffoli gates, and at the baseline's 1,457 qubits their two circuits need 27,519,242 and 28,047,551 Toffoli gates~\cite{initio2026}.
\cref{tab:published} compares the published designs whose counts we could check.
These designs minimize the qubits, the Toffoli gates or their product for one run, but we found none that minimizes the largest run under a budget for the total.

\begin{table*}[t]
\centering
\caption{Published designs for 256-bit curves and ours. Runs is the number of runs in one attempt at the key, a mean for Chevignard et al., and $B/B_0$ is the volume of one run relative to the baseline's. The counts of Babbush et al.\ are their stated upper bounds for a whole run, which they derive from the average executed Toffoli count of their addition.}
\label{tab:published}
\begin{tabular}{lrrrr}
\toprule
\textbf{Design} & \textbf{Qubits} & \textbf{Toffoli per run} & \textbf{Runs} & $\boldsymbol{B/B_0}$ \\
\midrule
H\"aner et al.~\cite{haner2026} & 1,457 & $3.93\times10^{7}$ & 1 & 1 \\
Babbush et al.~\cite{babbush2026} & $<1{,}200$ & $<9\times10^{7}$ & 1 & $<1.89$ \\
Babbush et al.~\cite{babbush2026} & $<1{,}450$ & $<7\times10^{7}$ & 1 & $<1.78$ \\
Luo et al.~\cite{luo2026} & 835 & $1.98\times10^{9}$ & 1 & 28.9 \\
Chevignard et al.~\cite{chevignard2026} & 1,193 & $5.42\times10^{11}$ & 22 & $1.13\times10^{4}$ \\
Initio Intelligence~\cite{initio2026} & 1,380 & $2.82\times10^{7}$ & 1 & 0.682 \\
Initio Intelligence~\cite{initio2026} & 1,457 & $2.75\times10^{7}$ & 1 & 0.701 \\
Initio Intelligence~\cite{initio2026} & 1,457 & $2.80\times10^{7}$ & 1 & 0.714 \\
Ours, unit-orbit & 1,246 & $1.26\times10^{7}$ & 79 & 0.275 \\
Ours, mixed-radix and root-sign & 1,246 & $1.19\times10^{7}$ & 102 & 0.259 \\
\bottomrule
\end{tabular}
\end{table*}

\paragraph{Splitting and tradeoffs.}
Eker{\aa} and H{\aa}stad shortened the runs for discrete logarithms known to be short~\cite{ekerahastad2017}, and Eker{\aa} extended the tradeoff to general discrete logarithms and orders, with lattice post-processing that recovers the logarithm from many runs~\cite{ekera2021}.
For groups of known order, such as elliptic-curve groups, Eker{\aa} analyzes a variant with about $(1+1/s)\,m$ group operations per run instead of $(1+2/s)\,m$ for an $m$-bit logarithm, so that a run keeps $256+\sigma+\lceil 256/s\rceil$ exponent bits at 256 bits~\cite{ekera2019revisiting}.
Gidney and Eker{\aa} build their estimate for factoring 2048-bit RSA integers on the algorithm of Eker{\aa} and H{\aa}stad~\cite{gidney2021rsa}.
Regev's factoring algorithm uses many runs of a multidimensional kind~\cite{regev2023}, and Ragavan and Vaikuntanathan made it space-efficient and robust to corrupted runs~\cite{ragavan2024}.
H\"aner et al.\ note that more instances of the algorithm could shrink each instance, and they leave this out~\cite{haner2026}.
These works give each run's saving by design, with heuristic or proven success for their decoders, but they give no lower bound on the size of a run.

\paragraph{Generic lower bounds.}
Shoup and Maurer bound the work of classical generic algorithms for discrete logarithms~\cite{shoup1997,maurer2005}, and \cref{sec:background} uses their bound to size the classical budget.
Hhan, Yamakawa and Yun work in a quantum generic group model and prove that a hybrid algorithm with $Q$ group operations in all must make $\Omega(\log|G|/\log Q)$ quantum group operations, of depth $\Omega(\log\log|G|-\log\log Q)$, between some two consecutive forced measurements~\cite{hhan2024}.
Hhan proves further bounds by compression arguments, for several instances of the problem and for order finding, and he treats algorithms that run a classical generic computation before their quantum part~\cite{hhan2024generic}.
These results bound what generic algorithms can achieve with given numbers of group operations or a given quantum depth.
\cref{thm:info} instead bounds the information of one linear run by its number of exponent values, which is what a floor on the run's exponent bits needs.

\paragraph{Endomorphisms and windows.}
Gallant, Lambert and Vanstone use an efficient endomorphism to halve the length of scalars in classical scalar multiplication~\cite{glv2001}, and Xu et al.\ cut the precomputation of curves such as secp256k1 by about five sixths with digit sets that the six units map to themselves~\cite{xu2024}.
In quantum circuits, signed windows use the unit $-1$ to halve the table~\cite{haner2020}, on top of windowed arithmetic~\cite{gidney2019windowed}.
Our unit-orbit windows extend quantum signed windows from the unit $-1$ to all six units of the curve, and they apply the table symmetry that classical scalar multiplication uses inside a quantum circuit whose cost we count.

\paragraph{Certificates for approximate arithmetic.}
Proos and Zalka start the accumulator at a random point so that exceptional additions hit a random fraction of inputs~\cite{proos2003}.
H\"aner et al.\ add a fresh random offset to every table and a fresh random mask to every approximate adder, which gives every addition uniform operands for every input, and they average the state over these choices to bound the success of one run~\cite{haner2026}.
Record-and-replay arithmetic bounds the error of each call by twice the square root of its bad mass and sums over the calls~\cite{long2026replay}, and approximate encoded permutations add up a deviation per addition in the same way~\cite{gidney2019approx}.
Gu et al.\ sum the failure probabilities of the components of point-addition circuits for secp256k1 that a search with large language models found, and they observe four failures in 100,000 tests of a complete circuit~\cite{gu2026}.
On the decoding side, Ragavan and Vaikuntanathan detect corrupted runs from their values by lattice reduction and drop them~\cite{ragavan2024}.
Our certificate takes the uniform operands and the averaging step from these works, and it adds a run-wide scalar that commutes with the units, a bound on every outcome of a run, flags that the circuit raises, and a success theorem for a decoder that drops the flagged runs.

\section{Conclusion}
\label{sec:conclusion}

In this work, we proved that one linear run reveals at most $\log_2(1+(N-1)/r)$ bits about the key, so that inside a total budget of $10\,T_0$ splitting alone at the baseline's arithmetic shrinks the largest run at most 1.86 times when the runs are fixed in advance.
Our construction takes one route past this limit.
We made each exponent bit cheaper with windows that use the six units of secp256k1, with a cheaper last addition and with a point-addition core borrowed from the ECDSA.Fail leaderboard, and we certified the success of the split runs for every key with a run-wide scalar, flags and a subset decoder.
Our best configuration shrinks the largest run 3.87 times at a total volume of $8.79\,T_0$, with 15 additions per run and 10.7 exponent bits above the floor at its run count.
Among the shorter configurations that we priced, none fits both budgets, and each of the six further levers that we priced outside the core is worth 0.3\% of $B$ or less, so cheaper point additions are the largest saving that our designs leave.

\begin{acknowledgments}
AI coding agents (Anthropic's Claude Opus 5.5 and Claude Sonnet 5.5) carried out the search and the circuit constructions under the authors' direction. A separate AI model (OpenAI's GPT-6 Astra and GPT-6.2 Astra) and authors have examined the results. The authors take full responsibilities.
X.G. acknowledges funding from the U.S. Department of Energy, Quantum Systems Accelerator.
\end{acknowledgments}

\appendix

\section{Proofs for Section 3}
\label{app:proofs}

This appendix proves \cref{thm:info}, \cref{cor:fano} and \cref{thm:adaptive}, extends \cref{thm:info} to runs that keep only a class of the final point, proves the floor inside a budget for runs of any sizes, and gives the exact computations on hash runs behind \cref{sec:limit-scope}.
Throughout, we identify a point with its discrete logarithm in $\mathbb{Z}_r$.

\subsection{Proof of Theorem 1}

Fix a linear run with set $S$, amplitudes $\alpha_z$ and generators $P_i = [u_i+v_i d]G$.
If the run uses randomness drawn independently of $d$, the ensemble keeps that randomness in a classical register whose law is the same for every key, so its Holevo information is the average, over the randomness, of the Holevo information for each fixed value.
We therefore fix the randomness, and then the set, the amplitudes, the numbers $u_i$ and $v_i$ and the measurement do not depend on $d$ (\cref{def:linear}).
After the classes of Step 0, we bound the Holevo information in six steps.

\paragraph{Step 0. Classes.}
Let $C_1,\dots,C_N$ be the classes of $S$ under the equivalence of \cref{sec:limit-thm}, let $|\phi_k\rangle = \sum_{z\in C_k}\alpha_z|z\rangle$, and let $\pi_k = \|\phi_k\|^2$.
All vectors of one class map to the same point for every key, which we write $f_d(C_k)$.
Two vectors $z\in C_k$ and $z'\in C_l$ of different classes, with $\Delta = z-z'$, map to the same point exactly when $(v\cdot\Delta)\,d \equiv -(u\cdot\Delta) \pmod r$.
If $v\cdot\Delta\not\equiv 0$, this equation has exactly one root because $r$ is prime, and if $v\cdot\Delta\equiv 0$, it has none because then $u\cdot\Delta\not\equiv 0$.
Every pair of classes therefore collides for at most one key.

\paragraph{Step 1. The reference state.}
After the run discards the group register, the exponent register holds
\[
\rho_d = \sum_{P} |\Phi_P\rangle\langle\Phi_P|, \qquad |\Phi_P\rangle = \sum_{k:\, f_d(C_k)=P} |\phi_k\rangle,
\]
and a measurement of the group register whose result is never used leaves the same state.
The reference state $\sigma = \sum_k |\phi_k\rangle\langle\phi_k|$ keeps the classes and drops every interference between them.
It does not depend on $d$, because the classes and the amplitudes do not.

\paragraph{Step 2. Relative entropy.}
For the ensemble $\{1/r,\rho_d\}$ with average state $\bar\rho$ and any state $\sigma$, $\mathbb{E}_d D(\rho_d\|\sigma) = \mathbb{E}_d D(\rho_d\|\bar\rho) + D(\bar\rho\|\sigma)$.
The Holevo information $\chi = \mathbb{E}_d D(\rho_d\|\bar\rho)$ is therefore at most $\mathbb{E}_d D(\rho_d\|\sigma)$, and the support of every $\rho_d$ lies inside that of $\sigma$.

\paragraph{Step 3. R\'enyi divergence.}
The relative entropy is at most the R\'enyi divergence of order two, $D(\rho\|\sigma)\le\ln\mathrm{Tr}[\rho^2\sigma^{-1}]$ with $\sigma^{-1}$ taken on the support of $\sigma$, because the Petz R\'enyi divergences do not decrease with their order~\cite{tomamichel2016}.

\paragraph{Step 4. The trace.}
We first drop every class of weight $\pi_k = 0$, which changes neither $\rho_d$ nor $\sigma$ and can only lower $N$.
Different classes hold disjoint sets of exponent vectors, so $|\phi_k\rangle = \sqrt{\pi_k}\,|\hat\phi_k\rangle$ with orthonormal $|\hat\phi_k\rangle$, and $\sigma^{-1} = \sum_k \pi_k^{-1}|\hat\phi_k\rangle\langle\hat\phi_k|$ on its support.
We call the set $F$ of classes that share one point at $d$ a fibre.
Its vector has $\|\Phi_F\|^2 = \sum_{k\in F}\pi_k$ and $\langle\Phi_F|\sigma^{-1}|\Phi_F\rangle = |F|$, and the vectors of different fibres are orthogonal.
Hence
\begin{align*}
\mathrm{Tr}[\rho_d^2\sigma^{-1}] &= \sum_F \|\Phi_F\|^2\,\langle\Phi_F|\sigma^{-1}|\Phi_F\rangle\\
&= \sum_k \pi_k\,n_k(d) = 1+x_d,\\
x_d &= \sum_k \pi_k\,(n_k(d)-1),
\end{align*}
where $n_k(d)$ is the size of the fibre of class $k$ at $d$.

\paragraph{Step 5. Average over the key.}
The number $n_k(d)-1$ counts the classes that collide with class $k$ at $d$, and by Step 0 each pair of classes collides for at most one of the $r$ keys.
For a uniform key, $\mathbb{E}_d\,x_d \le \sum_k \pi_k (N-1)/r = (N-1)/r$.

\paragraph{Step 6. Jensen's inequality.}
Steps 2 to 4 give $\chi\le\mathbb{E}_d\ln(1+x_d)$, and the concavity of the logarithm with Step 5 gives $\chi\le\ln(1+\mathbb{E}_d\,x_d)\le\ln(1+(N-1)/r)$.
The Holevo information bounds the information that any measurement of the exponent register reveals~\cite{holevo1973}, and $N\le|S|$, which proves \cref{thm:info}.

A script evaluated each link of this chain, $\chi\le\mathbb{E}_d D(\rho_d\|\sigma)\le\ln(1+\mathbb{E}_d\,x_d)\le\ln(1+(N-1)/r)$, with explicit density matrices on 300 random instances with complex amplitudes.
Every link held on every instance, and the largest ratio of $\chi$ to the bound was 0.72.

\subsection{Proof of Corollary 1}

Given $d$, the runs are independent, so their joint state is $\rho_d^{(1)}\otimes\cdots\otimes\rho_d^{(R)}$, and the product of their reference states does not depend on $d$.
Relative entropy adds over tensor products, so Steps 2 to 6, applied run by run, bound the Holevo information of the joint ensemble by $\sum_t\ln(1+(N_t-1)/r)$.
This bound covers every joint measurement of the runs, including one that chooses later measurements from earlier outcomes, because the runs themselves are fixed in advance.
Let the post-processing output a list $L$ of $K$ candidates, and let $P_e = \Pr[d\notin L]$.
Fano's inequality with a list gives $H(d\mid\text{outcomes})\le h(P_e)+(1-P_e)\log_2 K+P_e\log_2 r$, where $h$ is the binary entropy.
With $H(d\mid\text{outcomes}) = \log_2 r - I$ and $h(P_e)\le 1$, this gives $(1-P_e)(\log_2 r-\log_2 K)\le I+1$, which is \cref{cor:fano} with the information $I$ in bits.

\subsection{Outputs that are classes}

A run may keep only a class of its final point, such as the $x$-coordinate, which identifies a point with its negative.
\cref{thm:info} extends to such runs with a small factor.

\begin{corollary}
\label{cor:class}
Let the group register of a linear run hold only the class of the final point under a finite group $\Gamma$ of maps $s\mapsto\zeta s+l$ of $\mathbb{Z}_r$.
Then \cref{thm:info} holds with $(N-1)/r$ replaced by $|\Gamma|(N-1)/r$.
\end{corollary}

\begin{proof}
Two classes now collide when $f_d(C_l) = \gamma(f_d(C_k))$ for some $\gamma\in\Gamma$.
Each $\gamma$ gives one affine equation in $d$, so a pair of classes collides for at most $|\Gamma|$ keys, once the relations that hold for every key are merged into classes as in Step 0.
Step 5 gains the factor $|\Gamma|$, and the other steps are unchanged.
\end{proof}

The $x$-coordinate has $|\Gamma| = 2$, and the cube of the $x$-coordinate on a curve with $j$-invariant 0 has $|\Gamma| = 6$.
These class outputs therefore lower the floor by at most $\log_2 6 < 2.6$ exponent bits per run, a sixth of a window of 16 bits.
On the small groups that we computed exactly, class outputs never carried more information than the full point at the same number of classes.

\subsection{Proof of Theorem 2}

Fix $d$, a run $t$ and a history $h$ of earlier outcomes and randomness.
By \cref{def:linear}, the history fixes the run's set, amplitudes, numbers $u_i$ and $v_i$ and measurement, and none of them depends on $d$.
Let $\rho_d(h)$ and $\sigma(h)$ be the run's state and its reference state from Step 1, so that $\sigma(h)$ does not depend on $d$ either.
Let $q_d(h)$ be the weight of the classes whose point another class shares at $d$.
Fibre by fibre, $\rho_d(h)-\sigma(h)$ is the difference of two positive operators whose traces both equal the fibre's weight, so $\|\rho_d(h)-\sigma(h)\|_1\le 2q_d(h)$.
No measurement therefore separates the run from its reference by more than $q_d(h)$ in total variation, and by Step 5, $\mathbb{E}_d\,q_d(h)\le(N_t(h)-1)/r$.

The proof replaces the runs by their references one at a time, as in the hybrid argument of Bennett et al.~\cite{bennett1997}.
Let $H_j$ be the process in which runs 1 to $j$ are replaced by their reference runs and runs $j+1$ to $R$ are real, each chosen by the same rule from its history.
The process $H_R$ does not depend on $d$, because the reference states do not and the rules see only outcomes and randomness.
The processes $H_{j-1}$ and $H_j$ reach run $j$ with the same history distribution, differ only in run $j$ and treat the later runs alike, so their outcome laws differ by at most $\mathbb{E}_h\,q_d^{(j)}(h)$ in total variation, with $h$ drawn from the reference histories.
For every key, $\Pr[d\in L]\le\Pr_{H_R}[d\in L]+\sum_j\mathbb{E}_h\,q_d^{(j)}(h)$.
For a uniform key, the first term is at most $K/r$, because the list of $H_R$ has $K$ entries and does not depend on $d$.
The reference histories do not depend on $d$ either, so the averages over $d$ and over $h$ commute, and the second term is at most $\sum_j\mathbb{E}_h[N_j(h)-1]/r\le R(N-1)/r$.

\subsection{The floor inside a budget for runs of any sizes}
\label{app:frontier}

\cref{sec:limit-floor} states the floor inside a budget for runs of any sizes, and this section proves it.
Let $v$ be the baseline's volume per window of 16 exponent bits.
In the cost model of \cref{sec:limit-model}, a run of $D \ge 16$ exponent bits has volume $y = (D/16-1)\,v$, and \cref{thm:info} credits it with at most $f(y) = \log_2(1+(2^D-1)/r)$ bits.
A run of at most 16 bits has volume zero and is credited with at most $f(0) = \log_2(1+(2^{16}-1)/r)$, which is positive but less than $2^{-239}$ bits.
The exponent bits $D$ are affine in $y$, and the derivative $2^D/(2^D+r-1)$ of $\log_2(1+(2^D-1)/r)$ in $D$ rises with $D$, so $f$ is convex.

Fix a size $D$ for the largest run, a budget $V$ and a number $R$ of runs, and let $y_{\max} = (D/16-1)\,v$.
The total information $\sum_i f(y_i)$ is a convex function of the volumes $(y_1,\dots,y_R)$, so its largest value on the polytope $\{0\le y_i\le y_{\max},\ \sum_i y_i\le V\}$ is reached at a vertex.
At a vertex, at most one $y_i$ lies strictly between 0 and $y_{\max}$.
If $R\,y_{\max}\le V$, the largest value is therefore $R\,f(y_{\max})$, the value of $R$ equal runs.
Otherwise, with $k = \lfloor V/y_{\max}\rfloor$, it is $k\,f(y_{\max}) + f(V-k\,y_{\max}) + (R-k-1)\,f(0)$, the value of $k$ runs of the largest size, one run with the leftover volume and $R-k-1$ runs of volume zero.
The last term grows without limit in $R$, so the floor needs the bound $R\le 2^{40}$ of \cref{sec:limit-model}, under which the runs of volume zero add less than $2^{40}f(0) < 2^{-199}$ bits.
The smallest largest run is the smallest $D$ whose largest value reaches the 173.24 bits that \cref{cor:fano} requires.

At $V = 30\,B_0$, this smallest $D$ is 258.977 bits for every $R$ from 55 to $2^{40}$, with $k = 55$ runs at the largest size, so splitting alone shrinks the largest run by at most $1/0.5387 = 1.856$ times.
The leftover volume, $0.37\,B_0$, pays for a run of about 183 bits, which adds less than $2^{-72}$ bits, so the floor lies less than $2^{-78}$ bits below the 258.977 bits of 55 equal runs.
At $V = 12\,B_0$, the budget of $4\,T_0$, the same computation gives 21 runs of 264.245 bits and a factor of 1.817, and the leftover run of 215 bits adds less than $2^{-40}$ bits.
The proof of \cref{thm:adaptive} bounds the success by $K/r$ plus a sum of $(N_t-1)/r$ over the runs, and each term is also convex in the volume of its run.
For schedules whose every history stays within both budgets, the same argument therefore gives 57 runs of 250.153 bits at $V = 30\,B_0$ and a factor of 1.926, and the leftover run and the runs of volume zero add less than $2^{-56}$ to the sum.

\subsection{Runs that keep a hash of the point}

A hash run prepares $\sum_z\alpha_z|z\rangle|H(f_d(z))\rangle$ for a function $H$ of the point and uncomputes the point, so it pays for computing the point and for uncomputing it.
For a hash of $b$ bits known to the post-processing, the Holevo information of exact runs on small groups exceeds the bound of \cref{thm:info} 2.4 to 5.3 times (\cref{tab:hash}).
The same runs with a random hash unknown to the post-processing, averaged over the hash, stay below the bound, so the excess comes from knowing the hash.
Hash runs lie outside \cref{def:linear}, because the step that would limit them is an argument.
It concerns the direct test of a candidate key against a run, which evaluates the hash at the run's $|S|$ points at one group operation each in the generic group model.

\begin{table*}[t]
\centering
\caption{Holevo information in bits of one hash run on small groups, with the hash known to the post-processing, against the bound of \cref{thm:info}. Each entry is the mean over 8 random hashes of $b$ bits, $c = N/r$, and the factor over the bound is in parentheses.}
\label{tab:hash}
\begin{tabular}{rrrrrr}
\toprule
$r$ & \textbf{Box} ($c$) & \textbf{Bound} & $b=1$ & $b=2$ & $b=4$ \\
\midrule
101 & $8\times8$ (0.63) & 0.699 & 2.59 (3.7) & 3.02 (4.3) & 2.09 (3.0) \\
211 & $12\times12$ (0.68) & 0.747 & 3.21 (4.3) & 3.93 (5.3) & 3.06 (4.1) \\
211 & $20\times20$ (1.90) & 1.532 & 3.61 (2.4) & 4.87 (3.2) & 4.36 (2.8) \\
\bottomrule
\end{tabular}
\end{table*}

\section{Proofs for the certificate}
\label{app:cert}

This appendix proves \cref{lem:outcome}, couples each run with an ideal run, and proves \cref{thm:decoder}.

\paragraph{Setting.}
Let $x$ range over the set $X$ of inputs that the preparation of a run's exponent registers reaches, with amplitudes $a_x$, and let $D = \log_2|X|$ be the run's exponent bits.
The mixed-radix runs have $D = 267.88$ on 270 qubits, because each seven-level register takes 7 of its 8 values.
Let $q(o,x)$ be the amplitude with which the measurement maps input $x$ to outcome $o$.
The measurement is a Fourier transform over the $|X|$ values that the registers take, so $|q(o,x)| = |X|^{-1/2} = 2^{-D/2}$.
Each run draws its start $R_0$ uniformly from the group and its scalar $c$ uniformly from $[1,r-1]$, and each addition draws its own uniform measurement record.
On the ideal path of input $x$, the group register ends at $R_0+[c\,L(x)]G$ for the ideal exponent value $L(x)$.
The scalar only relabels the characters of the group register, so we index them by the scaled generator $[c]G$.
The character state $u$ then gives the point $[cs]G$ the amplitude $\chi_u([cs]G) = r^{-1/2}\omega^{us}$, with $\omega = e^{2\pi i/r}$, and neither the ideal law of $(o,u)$ nor the outcome law of the exponent registers depends on $c$.
Let $N_{\mathrm{DO}}(x)$ and $N_{\mathrm{P}}(x)$ count the additions on $x$'s ideal path that fail in the two ways that define $f_{\mathrm{DO},w}$ and $f_{\mathrm{P},w}$ in \cref{sec:certificate}.
We say that $x$ recovers if $N_{\mathrm{DO}}(x)\ge1$ and its branch still ends at the right point, clean and unflagged, and we set
\begin{multline*}
\mathrm{corr}_x = \bigl(N_{\mathrm{DO}}+2N_{\mathrm{P}}-1+(-1)^{N_{\mathrm{P}}}\,[N_{\mathrm{DO}}=0]\bigr)\\
\times[N_{\mathrm{DO}}+N_{\mathrm{P}}\ge2],
\end{multline*}
which is zero on paths with at most one failure and nonnegative on the others.
The second-order remainder of \cref{lem:outcome} is
\begin{multline*}
e_0 = \sum_x |a_x|\,2^{-D/2}\Bigl(\mathbb{E}[\mathrm{corr}_x]+\Pr[x\text{ recovers}]\\
+\Pr[x\text{ ends at a wrong point on the curve}]\Bigr).
\end{multline*}

\subsection{Proof of Lemma 1}

\paragraph{(i) Uniform operands.}
Fix an input $x$.
On its ideal path, addition $w$ has the operands $A_w = R_0+[c]\,S_w(x)$ and $B_w = [c]\,T_w(x)$.
For each $c$, $A_w$ is uniform because $R_0$ is, so $A_w$ is independent of $c$, and $B_w$ is uniform on the group without its identity because $T_w(x)$ is not the identity and $r$ is prime.
With its own uniform record, addition $w$ therefore fails in the first way with probability $f_{\mathrm{DO},w}$ and in the second way with probability $f_{\mathrm{P},w}$, for every $x$, so $\mathbb{E}[N_{\mathrm{DO}}(x)] = \sum_w f_{\mathrm{DO},w}$ and $\mathbb{E}[N_{\mathrm{P}}(x)] = \sum_w f_{\mathrm{P},w}$.

\paragraph{(ii) The amplitude of an outcome.}
Fix the start $R_0$, the scalar and the records $m$ of all additions, and consider the event in which no flag is raised, the ancillas end clean, and the run shows outcome $o$ and character label $u$, a part of the event of the lemma, so that a lower bound on it suffices.
By the window interface, the branch of each input is a basis state times a phase, so the event has amplitude $A = \sum_x a_x\,q(o,x)\,\beta_x$, where $\beta_x$ is the amplitude with which the branch of $x$ ends in the event.
The interface also gives every branch the same modulus $|c_m|$, whose square is the probability of the records $m$ for every input, and every good branch the same factor $c_m$, the phase of the records included (\cref{app:a2g}).
If the branch ends at the right point $P_x = R_0+[c\,L(x)]G$, clean and unflagged, then $\beta_x = \gamma_x\,c_m\,\chi_u(P_x)$, where $\chi_u(P)$ is the amplitude of the point $P$ in the character state $u$, of modulus $r^{-1/2}$, and $\gamma_x$ is the phase of the branch relative to the good branches.
If the branch is flagged, dirty or off the curve, then $\beta_x = 0$, because such a branch is orthogonal to every character state of the group with clean ancillas and no flag.
If the branch ends clean and unflagged at a wrong point on the curve, then $|\beta_x|\le r^{-1/2}|c_m|$.
Since $\chi_u(R_0+P) = r^{1/2}\chi_u(R_0)\,\chi_u(P)$, every good branch shares the factor $\eta = r^{1/2}c_m\,\chi_u(R_0)$, of modulus $|c_m|$.
We divide by it and let $\hat A = A/\eta$, which removes the factor of the records and the phase of the start point.

\paragraph{(iii) The phase of a good branch.}
If $N_{\mathrm{DO}}(x) = 0$, the branch follows its ideal path to the end, because an addition that fails only in its phase keeps the point and the ancillas, so $\gamma_x = (-1)^{N_{\mathrm{P}}(x)}$.
If $N_{\mathrm{DO}}(x)\ge1$, the branch ends at the right point only if $x$ recovers.
Hence
\begin{align*}
\gamma_x\,[x\text{ ends good}] &= 1-N_{\mathrm{DO}}(x)-2N_{\mathrm{P}}(x)\\
&\quad+\mathrm{corr}_x+\mathrm{rec}_x,\\
|\mathrm{rec}_x| &\le [x\text{ recovers}].
\end{align*}
For paths with at most one failure this is an identity with $\mathrm{corr}_x = 0$, and for paths with two or more failures it defines $\mathrm{corr}_x$.

\paragraph{(iv) The average over the run's randomness.}
By (i) and (iii), $\mathbb{E}[\gamma_x\,[x\text{ ends good}]] = g_0+\mathbb{E}[\mathrm{corr}_x]+\mathbb{E}[\mathrm{rec}_x]$ for every $x$, with $g_0 = 1-\sum_w(f_{\mathrm{DO},w}+2f_{\mathrm{P},w})$.
Averaged over the start, the scalar and the records, $\hat A$ is therefore $g_0$ times the ideal amplitude $A_I(o,u)$ plus terms from $\mathrm{corr}_x$, $\mathrm{rec}_x$ and the wrong points on the curve.
These terms have modulus at most $r^{-1/2}e_0$, because $|a_x\,q(o,x)| = |a_x|\,2^{-D/2}$ and $|\beta_x/\eta|\le r^{-1/2}$ for every branch.

\paragraph{(v) Jensen's inequality.}
The event has probability $|A|^2 = |c_m|^2\,|\hat A|^2$, and $|c_m|^2$ is the probability of the records.
Summed over the records and averaged over the start and the scalar, the probability of the event is therefore the average of $|\hat A|^2$ over the start, the scalar and uniform records, and Jensen's inequality gives $\mathbb{E}|\hat A|^2\ge|\mathbb{E}\hat A|^2$.
The ideal probability of $(o,u)$ is $|A_I(o,u)|^2 = \kappa(o,u)^2/r$, so $|\mathbb{E}\hat A|\ge r^{-1/2}(g_0\,\kappa(o,u)-e_0)$ whenever the right side is positive, which proves \cref{lem:outcome}.

\paragraph{(vi) A last addition that measures its end point.}
The root-sign last addition measures the end point itself, so no group register is left for the analysis to measure, and the label $u$ instead indexes the terms of a decomposition.
Fix the start, the scalar and the records, and let $A(P)$ be the amplitude of the event in which no flag is raised, the ancillas end clean, the run shows outcome $o$ and the measured end point is $P$.
By the window interface, $A(P) = \sum_x a_x\,q(o,x)\,\beta_x(P)$ (\cref{app:a2g}).
Here $\beta_x(P) = \gamma_x\,c_m\,[P = P_x]$ on a good branch and $\beta_x(P) = 0$ on a branch that is flagged, dirty or off the curve.
On a branch that ends unflagged at a wrong point $P'_x$, $\beta_x(P)$ vanishes for $P\ne P'_x$ and has modulus at most $|c_m|$ at $P'_x$.
Let $A(u) = \sum_P \chi_u(P)\,A(P)$.
On every branch, $\sum_P \chi_u(P)\,\beta_x(P)$ equals the $\beta_x$ of (ii) or obeys its bound, so $\hat A(u) = A(u)/\eta$ satisfies (iii) to (v) term by term, and the average of $|\hat A(u)|^2$ is at least $(g_0\,\kappa(o,u)-e_0)_+^2/r$.
The character states form an orthonormal basis of the $r$ points, so Parseval's identity gives $\sum_u |A(u)|^2 = \sum_P |A(P)|^2$, the probability of no flag, clean ancillas, outcome $o$ and the records $m$.
As in (v), the averages of $|\hat A(u)|^2$ over the start, the scalar and uniform records therefore add up over $u$ to $\Pr[\text{no flag}, \text{clean}, o]$, which is at most $\Pr[\text{no flag}, o]$, and each of them is nonnegative.
\cref{lem:coupling} uses \cref{lem:outcome} only through this sum, so it holds for this run, with $\Pr[\text{no flag}, o, u]$ read as the average of $|\hat A(u)|^2$.

The constants are the same for every key shift, because only the exact first lookup reads tables built from the public key.

\subsection{Good, flagged and bad runs}

\cref{lem:outcome} bounds every outcome of a run from below by a share of its ideal probability, so each run can be coupled with an ideal run.

\begin{lemma}
\label{lem:coupling}
For each run, independently of the other runs, there is a joint construction of an ideal draw $(u,\tilde o)$ from the ideal law, a uniform $U\in[0,1]$, and the real run's flags and outcome with their exact law, such that the real run is unflagged with outcome $\tilde o$ whenever $U\le K(\tilde o,u)$, where $K(o,u) = (g_0-e_0/\kappa(o,u))_+^2$.
\end{lemma}

\begin{proof}
Let $M(o) = \sum_u P_I(o,u)\,K(o,u) = \sum_u (g_0\,\kappa(o,u)-e_0)_+^2/r$.
Summing \cref{lem:outcome} over $u$ gives $M(o)\le\Pr[\text{no flag},o]$.
On the event $U\le K(\tilde o,u)$, the construction outputs $\tilde o$ without a flag, and otherwise it draws from the residual law $(\Pr[\cdot]-M)/(1-\sum_o M(o))$, which is nonnegative.
The output then has the exact law of the real run.
\end{proof}

We call a run good on the event of \cref{lem:coupling}, flagged if it shows a flag, and bad otherwise.
A good run's outcome follows the reweighted law $P_I(o,u)\,K(o,u)/(1-\Delta_0)$ with $\Delta_0 = 1-\mathbb{E}_I[K]$, and this law does not depend on the other runs.
Branches of different inputs are orthogonal, and the flags and the accumulator are measured in the computational basis, so a run is flagged with probability $\sum_x|a_x|^2\Pr[\text{the branch of }x\text{ ends flagged}]$, which the Monte Carlo bounds from below by $\varphi_{\mathrm{LB}}$.

\subsection{Proof of Theorem 3}

The labels good, flagged and bad are independent across runs, because each run draws its own randomness.
Turning a bad run into a flagged one never hurts the decoder.
With $b$ bad runs among $R'$ unflagged runs, a random subset of NEED runs avoids them with probability $\binom{R'-b}{\mathrm{NEED}}/\binom{R'}{\mathrm{NEED}}$, and flagging one bad run raises this to $\binom{R'-b}{\mathrm{NEED}}/\binom{R'-1}{\mathrm{NEED}}$.
The worst case therefore flags each run with probability exactly $\varphi_{\mathrm{LB}}$, leaves it bad and unflagged with probability $\nu = \Delta_0-\varphi_{\mathrm{LB}}$ and good with probability $1-\Delta_0$.
The numbers $n_F$ of flagged and $n_N$ of bad unflagged runs then follow the multinomial law $\mathrm{Mult}(n_F,n_N;R,\varphi_{\mathrm{LB}},\nu)$.

The decoder fails only if no attempt picks a subset without bad runs, or if the first such subset fails to decode.
The first event has probability $q(n_N,R-n_F)$ given $n_F$ and $n_N$, because the decoder's order depends only on the flags and on its own coins.
Here $q(b,R') = \binom{M-A_b}{T}/\binom{M}{T}$ with $M = \binom{R'}{\mathrm{NEED}}$, $A_b = \binom{R'-b}{\mathrm{NEED}}$ and $T = \min(\mathrm{TRIES},M)$, since the attempts use distinct random subsets.
Flagging a bad run keeps $A_b$ and lowers $M$, and $q$ can only fall when it does.
Each factor $(M-A_b-i)/(M-i)$ of $q$ grows with $M$, and once $M\le\mathrm{TRIES}$ the decoder tries every subset and misses them all only if $A_b = 0$.
For the second event, fix the decoder's order and a subset $s$.
The subset $s$ is the first one tried without bad runs exactly when all its runs are good and every subset tried before it holds a bad run outside $s$.
The second condition depends only on the runs outside $s$, so it is independent of the outcomes in $s$, which follow the reweighted law.
The first subset without bad runs therefore fails with probability at most $f_{\mathrm{good}}$, which proves \cref{thm:decoder}.

\subsection{Why the trials bound the real decoder}

The trials of \cref{sec:method} run the decoder on the worst case of the proof above.
The harness flags each run with probability exactly $\varphi_{\mathrm{LB}}$, keeps it good with probability $1-\Delta_0$, and marks the rest bad, so that every attempt on a subset that holds a bad run fails.
The success that the trials measure bounds the real decoder only if the real decoder fails at most as often as this harness decoder, and we prove this in three steps.

First, compare the real decoder with a poisoned decoder that sees the same flags, outcomes and coins but fails every attempt on a subset that holds a bad run.
Both try the same subsets in the same order, and an attempt on a subset of good runs gives the same result in both.
The real decoder therefore returns the key whenever the poisoned one does, at the same attempt or earlier, and an earlier answer is the key because every answer is checked against the public key.

Second, let $D$ be the number of subsets of good runs that decode to the key, which the outcomes of the good runs fix.
Given the labels and these outcomes, the poisoned decoder fails exactly when none of its $T$ attempts is one of these $D$ subsets, which has probability $\binom{M-D}{T}/\binom{M}{T}$.
Flagging a bad run keeps the good runs, their outcomes and $D$, and it lowers $M$, so this probability can only fall, as for $A_b$ in the proof above.

Third, a real run is good with the same probability $1-\Delta_0$ as a run of the harness, with the same law of its outcome, and it is flagged with probability at least $\varphi_{\mathrm{LB}}$.
The real labels thus arise from the harness's labels by flagging some of its bad runs, so the poisoned decoder fails at most as often on the real labels as on the harness's labels, where it is the harness's decoder.
The real decoder therefore fails with probability at most that of the harness, and the success bound of \cref{sec:method} holds for it.

\subsection{A toy check and the constants}

\paragraph{Toy check of the run-wide scalar.}
We computed both sides of \cref{lem:outcome} exactly on the group $\mathbb{Z}_r$, for $r = 31$ with two approximate additions and for $r = 37$ with three.
Each addition is a window of two qubits with signed odd digits behind an exact first lookup, and it fails according to a fixed table on its operands and record.
The tables are adversarial in three ways, with failures tied to the addend, failures tied to the accumulator, and mixtures that depend on the record.
Over every start, scalar and record, for three key shifts, every input failed each addition at the same mean rate, as step (i) says, and the bound held at every outcome and character label.
The smallest margin was between 0 and $8\times10^{-6}$ at $r = 31$ and between 0 and $1.2\times10^{-6}$ at $r = 37$.
Without the scalar, with $c = 1$ and failures tied to the addend, the failure rate of one input ranged from 0 to 0.32 at $r = 31$, and the bound with the constants of uniform operands failed by $4.7\times10^{-6}$.
The toy fails 5 to 30\% of its additions, far more often than the core, so it tests the inequality, and the constants at 256 bits come from the Monte Carlo of the next paragraph.

\paragraph{Constants and their confidence.}
The Monte Carlo of whole paths samples uniform operands and records along each configuration's own window sequence, with every addition at its abort cap, and each constant below rests on one-sided bounds at confidence $1-0.001$ each, five in all, so that the constants hold together at confidence $1-0.005$.
For the unit-orbit configuration, 262,144 paths of 15 additions bound $1-g_0$ by 0.04284, so $g_0 \ge 0.95716$, the path term $\mathbb{E}[\mathrm{corr}_x]+\Pr[x\text{ recovers}]+\Pr[\text{wrong point on the curve}]$ by $2.10\times10^{-3}$, and $\varphi_{\mathrm{LB}}$ from below by 0.03145.
For the mixed-radix configuration with a full last window, 393,216 paths give $g_0 \ge 0.95711$, a path term of at most $1.64\times10^{-3}$ and $\varphi_{\mathrm{LB}} \ge 0.03256$.
For the configuration with the root-sign last addition, 524,288 paths that replay that addition give $g_0 \ge 0.95930$, a path term of at most $1.41\times10^{-3}$ and $\varphi_{\mathrm{LB}} \ge 0.03115$, each rounded outward from the value that the harness uses at full precision.
We fixed the partition of the paths into blocks that the bounds on $1-g_0$ and on $\mathbb{E}[\mathrm{corr}_x]$ use after the batches had run, so we also bound both means without it.
The number of batches did not depend on their results.
Each batch holds 131,072 paths, a size set by the memory of one process, and the unit-orbit configuration splits its planned 262,144 paths into two batches.
The mixed-radix configuration with a full last window uses the three batches that had finished when its trials were registered, and batches that finished later did not enter its constants.
The root-sign configuration uses four batches that one job ran in sequence, with their number fixed at launch.
The bounds therefore hold at their stated confidence for a fixed number of paths.
Every path of $J = 15$ additions has values in $[0,2J]$, and the product $\prod_i(1-X_i/2J)$ over $n$ paths has expectation $(1-\mathbb{E}X/2J)^n$.
Markov's inequality therefore gives $\mathbb{E}X\le 2J\,[1-(\alpha\prod_i(1-X_i/2J))^{1/n}]$ at confidence $1-\alpha$, with the same $\alpha = 0.001$ as the block bound.
On the saved paths this bound lies below the block bound in all three configurations, by at least $6.0\times10^{-4}$ for $1-g_0$ and $2.7\times10^{-4}$ for $\mathbb{E}[\mathrm{corr}_x]$, so the constants hold without the partition.
The per-addition rates of these paths come from the classical replay of the core's gate stream, whose failure rate on the windowed core was $2.47\times10^{-3}$ over 327,680 inputs against the circuit checker's $2.21\times10^{-3}$ over 98,304 inputs (\cref{sec:method}), so the success bound is a bound from sampling.
A rerun of every batch of these paths with the semantics of the counted circuits gives the same counts and therefore the same constants (\cref{app:a2g-replay}).

The terms of $e_0$ depend only on the digits of the approximate windows.
In our configurations no approximate window reads a shaped qubit, because the register $b$ and the shaped top window belong to the exact first lookup.
The sum over inputs in $e_0$ therefore factors into the path term, measured on uniform inputs, times the product $\Lambda$ of the shaped registers' $L_1$ factors $\sum_x |a_x|/2^{k/2}$, where $a$ is the normalized amplitude vector of a shaped register over its $2^k$ values.
A uniform register, binary or seven-level, contributes a factor of 1, because $D$ counts only the values that a register takes.
Each $L_1$ factor is about 0.90, so $\Lambda = 0.736$ in the unit-orbit configuration, with an 11-bit register $b$ and a top window of 4 by 3 bits, and $\Lambda = 0.732$ in the mixed-radix ones, with a 9-bit register $b$ and a top window of 4 by 4 bits.
The harness thus uses $e_0 \le 1.55\times10^{-3}$, $1.20\times10^{-3}$ and $1.04\times10^{-3}$ in the three configurations.
For the root-sign configuration, a planning model of \cref{thm:decoder} with these constants and an assumed decoding rate of a clean subset put the failure probability below $3\times10^{-4}$ at 102 runs, which fixed the run count before the trials.

\section{The window interface of each circuit}
\label{app:a2g}

\cref{lem:outcome} assumes that every approximate addition of a run has the window interface of \cref{def:interface}.
This appendix derives that interface for each window type from its gates, on every branch of every input.
We first state the gate model and the interface at the level of gates.
Then, we prove the interface for each window type and list the checks that ran.
Finally, we describe the replays that produced the constants of \cref{app:cert} and their rerun with the semantics of the counted circuits.

\subsection{Gates, records and the interface}

Every window is a list of gates of the reversible-circuit checker.
The gates are $X$, CNOT, Toffoli, a Toffoli gate into a fresh qubit, $Z$, CZ, CCZ, SWAP and a global sign, each optionally under a classical condition.
Three further operations end the use of a qubit.
A measurement in the $X$ basis stores its outcome and releases its qubit, a flag measures its qubit in the $Z$ basis and reports a detected failure with the outcome 1, and a release frees a qubit.
We call the string of a window's outcomes in the $X$ basis its record $m$, and $k$, the number of these measurements, is fixed by the gate list.
The outcomes of the flags form a second record, which for the root-sign last addition also holds the measured end point.
For a fixed record $m$, let $K_m$ be the linear map from the window's input, made of the accumulator, the window register and fresh ancillas, to the unnormalized state after the window, in which every outcome in the $Z$ basis, the measured end point included, and every released wire stay as registers of the output.

\begin{definition}[Window interface at the level of gates]
\label{def:interface-gates}
Fix the tables of a run, and let window $w$ add the point $B = B_w(x_w)$ for the value $x_w$ of its window register.
Some measurements of the window in the $X$ basis unload a register, and we write $m_1,\dots,m_q$ for their outcomes.
The window has the interface if there are a sign $\theta(m)$ for each record and functions $\mathrm{out}$, $\mathrm{anc}$, $\mathrm{flag}$ and $\epsilon_0\in\{\pm1\}$ of $(A,B,m)$, the same for every table and scalar, such that the following four conditions hold for every basis input $|A\rangle|x_w\rangle|0\rangle$ and every record $m$.
\begin{itemize}
\item[(I1)] The window maps the input to one basis state times a sign,
\begin{multline*}
K_m|A,x_w,0\rangle = 2^{-k/2}\,\theta(m)\,\epsilon\\
\times|\mathrm{out}(A,B,m),\,x_w,\,\mathrm{anc}(A,B,m)\rangle,
\end{multline*}
and the flags read $\mathrm{flag}(A,B,m)$.
The sign is
\[
\epsilon = \epsilon_0(A,B,m)\,(-1)^{\sum_{i=1}^{q}\langle m_i,\delta_i\rangle},
\]
where $\delta_i$ is a bit vector that is 0 exactly when the $i$th unloaded register holds its ideal value.
Whether it does is a function of $(A,B)$ and of the outcomes outside $m_1,\dots,m_q$, and nothing else in the window reads $m_1,\dots,m_q$.
\item[(I2)] Every record has probability $2^{-k}$ on every basis input.
\item[(I3)] A good branch, with $\mathrm{out} = A+B$, $\mathrm{anc} = 0$, no flag and $\epsilon = +1$, carries the sign $\theta(m)$ alone.
\item[(I4)] The number of Toffoli gates that the window executes is a function of $m$ alone.
\end{itemize}
\end{definition}

A branch that is good except for $\epsilon = -1$ is phase-only, and every other branch deviates.
Conditions (I1) to (I3) restate \cref{def:interface} at the level of gates, with the signs of the unloads kept apart.
The reason is that, for a fixed record, the sign of a $y$ unload whose wires hold a wrong value can also depend on the sign of the digit (\cref{lem:unload}).
Over a uniform record, however, that sign is uniform and independent of the other outcomes, of the basis state and of the digit, because $\delta_i$ has a bit equal to 1 and the outcome that multiplies it is uniform by (I2) and read nowhere else.
The law of the window's action over a uniform record therefore depends on the digit only through $B$, as \cref{def:interface} requires.
A branch with a wrong unload is classed by its sign like any other branch, so when the rest of the branch is good, it is good or phase-only with probability $1/2$ each.

Condition (I4) lets every branch of a record abort together, so an abort is a visible flag of the whole run.
Steps (i) and (iv) of the proof of \cref{lem:outcome} use the law of (I1) and (I2), addition by addition along a path, because they need the classes and the end state of input $x$ under uniform records to depend on $x$ only through the operands.
Step (ii) uses (I1) to (I3), because the factors $2^{-k/2}\theta(m)$ of the windows multiply to the factor $c_m$ that every good branch carries.
In every window below, $\theta(m) = +1$, the convention of the circuit checker and of the replays, which both count a branch with the phase $-1$ as phase-only.

\begin{lemma}[Gates]
\label{lem:gates}
Let every classical condition of a gate list be a function of earlier outcomes in the $X$ basis.
Then for every record $m$ and every basis input $z$, $K_m|z\rangle = 2^{-k/2}(-1)^{\phi(z,m)}|F(z,m)\rangle$ for functions $\phi$ and $F$, the flags are a function of $(z,m)$, every record has probability $2^{-k}$, and the number of executed Toffoli gates is a function of $m$.
\end{lemma}

\begin{proof}[Proof idea]
The $X$, CNOT, Toffoli and SWAP gates permute basis states, and the $Z$, CZ, CCZ and sign gates multiply them by $\pm1$.
A measurement in the $X$ basis of a qubit that holds $v$ gives the outcome $m$ with the operator $2^{-1/2}(-1)^{mv}$, so each outcome has probability $1/2$ on every basis state, and a flag projects onto the value of its qubit.
Once $m$ is fixed, every condition is a fixed bit, and induction over the list gives the lemma.
\end{proof}

\subsection{The window types}

\paragraph{The borrowed core.}
Inside a window, the core's stream reads the addend $B$ only through the values $x_B$, $(3x_B+\mathrm{OFF}) \bmod p$, $x_B+1$, the bits 236 to 254 of $x_B$, and $y_B$, where OFF is the stream's squaring offset.
We write $V(B)$ for these values.
The window loads them at the core's three $x$ sites and two $y$ sites with select-swap reads and erases them by measurements in the $X$ basis.
The first $y$ site opens the stream, and the second sits near its end.

\begin{lemma}[Core]
\label{lem:core}
Suppose that at each $x$ site the site register holds its part of $V(B)$ and serves only as a control, and that each $y$ load puts $y_B$ into the stream's $y$ wires.
Then for every record $m$ of the stream, every classical bit and every condition of the stream is a function of $m$, and the stream acts on the accumulator and its own wires as a basis map times $\pm1$ that depends only on $(A,V(B),m)$.
No later condition reads a flag, and the number of Toffoli gates that the stream executes is a function of $m$.
\end{lemma}

\begin{proof}[Proof idea]
Our port keeps the raw stream's addend bits at 0 and never writes them, so its classical bits change only by outcomes and by the stream's classical operations, under conditions built from earlier outcomes.
The 4,974 site operations of the site map are the only operations that read $B$.
At every site operation, the port replaces the addend bit in the condition by the site register, so an $X$ gate becomes a CNOT, a CNOT becomes a Toffoli gate and a $Z$ gate becomes a CZ, and the $y$ loads write $y_B$ into the wires.
Every other operation is the checker's gate of its kind under its condition, and \cref{lem:gates} gives the basis map, the records and the executed count.
An exact scan of the raw stream found that no flag writes a classical bit and that no Toffoli gate sits under a bit that depends on a flag.
The three CNOT gates that read the addend become unconditional Toffoli gates, and every Toffoli gate of the digit layer is unconditional, so the abort cap stops a window exactly when the stream's executed count passes 726,000.
\end{proof}

The hypothesis on the $y$ loads holds only if the $y$ wires hold 0 when a load starts.
Into wires that held a value $v\ne0$, the select-swap load would leave a value that depends on $v$ under one value of its selector and would move $v$ into the measured partner slot under the other.
The core reuses these wires before its second load, so the next lemma shows that they hold 0 on every branch.

\begin{lemma}[Zero before load]
\label{lem:zeroload}
In every window, on every branch of every basis input and every record, the 256 $y$ wires of each $y$ site hold 0 when the site's load starts.
\end{lemma}

\begin{proof}[Proof idea]
An exact scan of the raw stream counts as a write every operation that can change a wire's value.
The wires of the first $y$ site are never written before its load, which opens the stream, so they are fresh.
Each of the 256 wires of the second $y$ site is written before its load.
For 248 of them, the last write is a flag or a measurement in the $X$ basis under no condition, after which our port releases the wire, so the load meets a fresh qubit.
Each of the other 8 is written after its last release only inside one block of the stream under a single condition, by a Toffoli gate, a measurement in the $X$ basis and a flag, the flag last.
A measurement or flag under a condition first moves the wire's value into a fresh qubit under that condition.
On branches where the condition holds, the wire therefore ends at 0, and on the other branches none of the three operations acts, so the wire keeps its 0.
\end{proof}

An unload, in contrast, meets whatever the stream left in the $y$ wires, which on a failing branch need not be $y_B$.

\begin{lemma}[Unload mismatch]
\label{lem:unload}
At each $y$ unload, a branch whose $y$ wires hold $u$ gains the factor $(-1)^{\langle m_y,\delta\rangle}$, where $m_y$ is the string of the unload's outcomes, which nothing else in the window reads, and $\delta$ is a bit vector that is 0 exactly when $u = y_B$.
Over uniform outcomes, this factor is therefore $+1$ if $u = y_B$ and otherwise a uniform sign that is independent of the other outcomes, of the basis state and of the digit.
The same holds for the unload of the slope register in the root-sign last addition, with $\delta = 0$ exactly when the register holds the slope that the measured sign selects.
\end{lemma}

\begin{proof}[Proof idea]
At a $y$ unload, the layer applies the exact negation by the digit's sign $s$, which maps $u$ to $N_s(u) = (p-u) \bmod 2^{256}$ for $s = 1$ and to $u$ for $s = 0$, then measures the 256 wires in the $X$ basis and releases them.
The phase oracle removes $(-1)^{\langle m_y,y_T\rangle}$ for the $y$-coordinate $y_T$ of the table point that the digit reads, so the branch keeps the factor $(-1)^{\langle m_y,\delta\rangle}$ with $\delta = N_s(u)\oplus y_T$.
The map $N_s$ is a bijection with $N_s(y_B) = y_T$, so $\delta = 0$ exactly when $u = y_B$.
No condition reads an outcome of the layer, and the oracle reads $m_y$ only through this factor.
If $\delta\ne0$, fix a bit of $\delta$ equal to 1, and the uniform outcome that multiplies it makes the factor a uniform sign, whatever the other outcomes and the branch.
The root-sign last addition erases its slope register in the same way, with $\delta = \lambda\oplus\lambda_s$ for the register's value $\lambda$ and the slope $\lambda_s$ that the measured sign selects (\cref{lem:rootsign}).
\end{proof}

\paragraph{Unit-orbit windows.}
A unit-orbit window wraps the core in a digit layer.
The layer converts the residue $(a,b)$ of the window register into its orbit form $(x'',s,j)$ of \cref{app:windows}, reads the table point of $x''$ at each site and applies the unit $(-1)^s\omega^j$ to it with forward ripples whose carries it measures in the $X$ basis.
One deferred phase oracle removes the phases that all of the layer's measurements leave, and the layer converts the orbit form back to $(a,b)$ at the end.

\begin{lemma}[Unit-orbit windows]
\label{lem:unitorbit}
The unit-orbit windows of 8 and 9 bits have the interface, with $B_w = [c]\,T_w[x_w]$ and $\theta = +1$.
\end{lemma}

\begin{proof}[Proof idea]
Three facts about the digit layer carry the proof.
First, the reads address exactly $x''$, each $x$ site's register ends its read holding its part of $V(B)$, and each load leaves $y_B$ in the $y$ wires, which hold 0 before it by \cref{lem:zeroload}.
Second, every register that the layer measures in the $X$ basis holds a function of $(x'',s,j)$, except the $y$ wires at an unload, so the phase oracle, which recomputes these values, removes every phase that the measurements leave except the factors of \cref{lem:unload}.
Third, every ancilla of the layer returns to 0, and the window register returns to $(a,b)$.
With these facts, the stream sees exactly $V(B)$ at every site on every branch, so by \cref{lem:core} its action, flags, phase and executed count are functions of $(A,V(B),m)$, and the map from $B$ to $V(B)$ depends on neither the table nor the scalar.
Whether the $y$ wires hold $y_B$ at an unload is one of these functions, so \cref{lem:unload} gives the sign of (I1).
\cref{lem:gates} gives the uniform records of (I2), the convention above gives (I3), and \cref{lem:core} gives (I4).
The three facts are how the layer is written, and the checks of \cref{app:a2g-checks} examine them on every residue at 256 bits and on every operand at small widths.
\end{proof}

\paragraph{Mixed-radix windows.}
A mixed-radix window runs the same core with the digit layer of the residues modulo $(3+\omega)2^7$ or $(2-\omega)2^7$.
Its layer takes the unit of the orbit from the $\mathbb{F}_7$ part $c$ of the residue when $c\ne0$ and from the conversion of the unit-orbit windows when $c = 0$, reads the table point of the orbit's representative at each site, and removes the recorded phases with a phase oracle over the orbit form.

\begin{lemma}[Mixed-radix windows]
\label{lem:mixed}
The mixed-radix windows over $(3+\omega)2^7$ and $(2-\omega)2^7$ have the interface.
\end{lemma}

\begin{proof}[Proof idea]
The core's stream loop of these windows is the loop of the unit-orbit windows line for line, so \cref{lem:core,lem:zeroload,lem:unload} apply unchanged.
The three facts of the proof of \cref{lem:unitorbit} hold for every residue $(c,a,b)$ with $c$ in $[0,7)$, and the checks of \cref{app:a2g-checks} examine them on all $7\cdot4^7$ residues of each window type.
The value $c = 7$ is not an input, because the preparation of the seven-level register puts no amplitude on it (\cref{lem:registers}).
\end{proof}

\paragraph{The last additions.}
The last addition of the unit-orbit configuration and of the mixed-radix configuration with a full last window is the 9-bit window of \cref{lem:unitorbit}, and the run then measures the accumulator and tests the curve equation.
The root-sign last addition instead measures the end point inside the addition (\cref{sec:core}).

\begin{lemma}[Full last window]
\label{lem:lastwindow}
The full last window has the interface, and the test of the curve equation on the measured accumulator is a visible flag that costs nothing in \cref{lem:outcome}.
\end{lemma}

\begin{proof}[Proof idea]
The first claim is \cref{lem:unitorbit} at $W = 9$.
The curve test is the projective measurement onto the span of the $r$ point encodings and its complement, and the character states span exactly that space.
The test followed by the analysis measurement of \cref{lem:outcome} is therefore the measurement onto the $r$ character states and the space off the curve.
An end off the curve is orthogonal to every event with no flag and a character label, which is the case $\beta_x = 0$ of step (ii) in \cref{app:cert}.
\end{proof}

\begin{lemma}[Root-sign last addition]
\label{lem:rootsign}
The root-sign last addition has the interface, with the measured end point in place of the output.
The end point is a function of the branch, and the tables and phase corrections after its measurement depend on it.
\end{lemma}

\begin{proof}[Proof idea]
Its prefix runs the core's stream up to the square that yields the new $x$-coordinate $x_J$, with the digit layer of \cref{lem:unitorbit}, and stops before the second $y$ site.
The prefix holds one $y$ site, which loads into fresh wires (\cref{lem:zeroload}) and leaves the factor of \cref{lem:unload} at its unload.
The one change in the prefix is the site that reads $(3x_B+\mathrm{OFF}) \bmod p$, which reads $(2x_B+\mathrm{OFF}) \bmod p$ instead, so that the register that the tail measures ends holding a function of $x_J$ alone.
The checks of \cref{app:a2g-checks} cover the unit gadget at this multiplier.
The tail then works in four steps.
First, it measures that register and four sign wires in the $Z$ basis and decodes them, and an invalid pattern, a value $x_J^3+7$ that is not a square, or any live nonzero wire of the stream raises a flag.
The decoder accepts only the sign encoding that the stream's rule fixes, so on every unflagged branch the measured data are a function of $x_J$ alone.
Second, a select-swap read over the digit loads 32 bits of one slope and 16 bits of the other from a classical table that the run builds from $x_J$ and its two roots, and it loads 0 for both when $x_B = x_J$.
Third, exact chains of AND gates compare both slopes with the slope register, and anything other than exactly one match raises a flag.
Fourth, measurements in the $X$ basis erase the read buffers and the slope register, and their phases, which are functions of the digit, $x_J$, the sign and the record, go into the window's one deferred phase oracle.
A branch whose slope register holds the slope that its sign selects, 0 when $x_B = x_J$, therefore ends with the phase $+1$, and any other unflagged branch carries the uniform sign of \cref{lem:unload}.
Every outcome in the $X$ basis is uniform by \cref{lem:gates}, and the data measured in the $Z$ basis are a function of the branch, so (I1) to (I4) hold with the measured point as the output.
Step (vi) of \cref{app:cert} carries \cref{lem:outcome} over to this measured end point.
\end{proof}

\paragraph{The first lookup and the exponent registers.}
The first lookup is the only part of a run that reads tables built from the public key, and the exponent registers are prepared and measured outside the additions.

\begin{lemma}[First lookup]
\label{lem:firstlookup}
For every address $k$ and every record, the first lookup maps $|k\rangle|0\rangle$ to $|k\rangle|T[k]\rangle$ with no phase and no garbage, where $T[k] = [c]\,T_1[k]+R_0$ is the run's classical table.
It never fails, so it carries no failure weight.
\end{lemma}

\begin{proof}[Proof idea]
The read writes $T[2i]$ and $T[2i]\oplus T[2i+1]$ by unary iteration over the high address bits, adds the difference into the output under the lowest address bit, and erases the difference by measurements in the $X$ basis.
The phase that this erasure leaves depends only on the high address bits, and a recursive phase lookup removes it at once.
The circuit checker ran the lookup in exact mode on every lane, with random and forced outcomes, at 17 and 18 address bits with 512-bit payloads on 192 lanes each and at 4, 6, 8, 10 and 12 address bits, and it found no error.
A version without the phase fix-up fails the same check.
The lookup is exact for every table, so the tables built from the shifted public key leave the constants of \cref{lem:outcome} unchanged.
\end{proof}

\begin{lemma}[Exponent registers]
\label{lem:registers}
The preparations and Fourier measurements of the exponent registers act on those registers only, are exact, and give the outcome law that the harness samples.
\end{lemma}

\begin{proof}[Proof idea]
\cref{app:windows} gives the circuits of the seven-level register and of the digit registers that correct the phase of each window, with their checks.
The digit registers of the mixed-radix windows also matched their digit map at window sizes 2, 3, 4, 5 and 7, and the sine states were checked gate by gate for every measurement outcome.
These circuits touch neither the accumulator nor the window's arithmetic, so they enter the interface only through the window register, which \cref{lem:unitorbit,lem:mixed} return unchanged.
\end{proof}

\subsection{Checks that ran}
\label{app:a2g-checks}

Two checks examine the digit layers of \cref{lem:unitorbit,lem:mixed}.

\paragraph{Every residue at 256 bits.}
The first check runs the production build of each window type at $n = 256$ on all of its residues.
Its input is a synthetic stream of the real length, 12,368,253 operations, that holds the 4,974 real site operations at their real indices with their entries in the site map and nothing else of the arithmetic.
Each operation at an $x$ site writes a fresh output wire, and each $y$ site's wires are copied out between the load and the unload.
The table comes from the window's specification, the expected addend is computed from the digit independently of the circuit, and the outcomes are random on every lane, with all 0 and all 1 added on the first batch.
On every lane, the check requires that every output wire equals its site value of the addend, that the $y$ copies equal $y_B$, that the window register is restored with the phase $+1$, and that no flag is raised and no nonzero qubit is freed or left over.
All residues passed, 65,536 at $W = 8$, 262,144 at $W = 9$ and 114,688 for each of the two mixed-radix windows.
The layer's Toffoli count on this stream, plus the 3 Toffoli gates that the three site CNOT gates become and the stream's 761,147, reproduces the emitted count of each counted window exactly, 814,662 at $W = 8$, 946,884 at $W = 9$ and 847,736 for the mixed-radix windows.
The check therefore ran the layers of the counted circuits.
It uses one table, so it covers the address decoding of every residue and the arithmetic on that table's operands.

\paragraph{Every operand at small widths.}
The second check covers the arithmetic for every operand that can occur, at widths $n = 5$ to 10 with $p \equiv 1 \pmod 3$.
It runs the gadget that applies the unit for every $x$ in $[1,p)$, every unit power and the multipliers 1, 2 and 3.
It also runs the constant addition that forms $(3x+\mathrm{OFF}) \bmod p$ and the increment that forms $x+1$ for every value, and the negation of $y$ for every $y$ in $[1,p)$, each with random, all-0 and all-1 outcomes.
All 108 configurations were exact, with no wrong lane, the phase 0 after the recorded phases, no nonzero qubit freed and no garbage.
These gadgets repeat the same full-adder step at every bit, so the small widths exercise every code path, namely the lowest bit, an interior bit, the top bit, the extra carry bit and the comparator of the reduction.
At $x = 0$ the unit gadget returns $p$ for the unit power 2, but 0 is not an operand.
The reason is that 7 is not a square modulo $p$, so no point of secp256k1 has $x = 0$, and a unit or a factor of 1, 2 or 3 keeps $x$ nonzero.

\subsection{The replays and the circuits}
\label{app:a2g-replay}

The constants of \cref{app:cert} come from path replays, classical programs that run the core's gate stream on many pairs of operands at once.
Each replay draws the operands of every addition with the law of a run, from a uniform start, one scalar per path and the addend $[c]\,T_w(x)$ of a uniform digit, draws a uniform record for every addition, and returns the class of every addition and the end class of every path.
The replays never see the digit, so a replay stands for a circuit only if the law of the circuit's action depends on the digit through the addend alone, which is what (I1) and (I2) give.
The replays behind the constants ran the core's raw stream, in which the addend is a classical bit string, whereas the circuits run our port of the stream, in which the addend is a quantum register that the stream reads at its 4,974 site operations.

To take the constants from the circuits themselves, we reran the three Monte Carlos with the port's semantics.
The rerun keeps the stream's classical addend bits at 0, as the port does, and applies every site operation with its site value of the addend.
It loads each $y$ site in one step and stops if a wire is nonzero before the load, which \cref{lem:zeroload} rules out.
At each unload, it releases the wires and multiplies every lane whose wires do not hold $y_B$ by a fresh uniform sign from a separate generator, which is the law of \cref{lem:unload}.
Its flags are exactly the events that the circuit shows, namely the core's flags, an abort, the flags of the root-sign tail and an end off the curve.
In the root-sign last addition, the third site reads $(2x_B+\mathrm{OFF}) \bmod p$, and the tail follows the circuit step by step, the zero payload at $x_B = x_J$ included.
The rerun takes its site tables from the builder's own dispatch, which a separate check found equal to the tables of all three window builders.
By \cref{lem:core,lem:zeroload,lem:unload,lem:unitorbit,lem:mixed,lem:rootsign}, the rerun therefore draws the class and the end state of each addition with the law that the counted window gives them on the same operands.

The rerun ran inside the unchanged Monte Carlo drivers, with their path law, seeds, abort caps and pooling.
It covered the nine batches of 131,072 paths that produced the constants, two for the unit-orbit configuration, three for the mixed-radix configuration with a full last window and four for the root-sign configuration.
It draws the outcomes of the measurements exactly as the earlier replays do, so the two give equal records as long as their semantics agree.
All nine batches gave the same counts as before, over 1,179,648 paths with 17,694,720 additions on their ideal operands and 56,510 additions that continue paths after a failure.
The seven batches that also store arrays per path gave the same arrays.
No load met a nonzero wire, and no unload met wires that did not hold $y_B$, on any lane of any batch, the lanes after a failure included.
The root-sign tail accepted 523,690 of the 524,288 lanes on ideal operands and none of the 3,246 lanes after a failure, and no accepted output had $x_B = x_J$ or a slope other than the one its sign selects.
In one batch of each configuration, the earlier replay also ran beside the rerun on every call, and the two agreed on every lane.
Every bound behind $g_0$, the path term and $\varphi_{\mathrm{LB}}$ therefore comes out as the same floating-point number, so the constants of \cref{app:cert} are path statistics of the counted circuits.
The end-to-end harness reads the batches only through these constants, and each trial draws its randomness from the configuration's seed and the trial's index.
The 563 trials of each configuration at 256 bits, their success bound of 0.994506, and the run counts and volumes of \cref{sec:eval} therefore stand.
A rerun of the first 20 of these trials for each configuration, from its frozen code and on the new batches, reproduced every one of them in outcome, run count, classical work and cost.

As a check at the level of gates, the circuit checker also ran its own Monte Carlo of the counted 9-bit window at 256 bits.
It ran the window on 32,768 inputs with a uniform accumulator and a uniform window value, in 16 batches of 2,048, one for each window position with that position's fixed table.
The window flagged 66 of these inputs, against 58.5 that the replay predicts from its average flag rate of $1.785\times10^{-3}$ per addition in the mixed-radix configuration with a full last window.
A Poisson count with mean 58.5 reaches 66 or more with probability 0.18.
The window failed on 84 inputs, flagged or not, against 81.5 from the replay's failure rate of $2.488\times10^{-3}$.
The gate-level counts thus agree with the replay's predictions, while $\varphi_{\mathrm{LB}}$ itself comes from the whole-path reruns above.

\section{Windows, circuits and counts}
\label{app:windows}

\paragraph{Unit-orbit digits.}
A residue $x = a+b\omega$ of $\mathbb{Z}[\omega]/2^W$ has a valuation $v$, the number of factors of 2 that divide it, and a unit power $j$ that the bit pair $(a_v,b_v)$ determines.
The circuit sets $x' = \omega^{-j}x$, reads a sign $s$ from bit $v+1$ of the first coordinate of $x'$, and takes $x'' = (-1)^s x'$ as the canonical representative of the orbit.
The window's digit is $(-1)^s\omega^j\,\mathrm{lift}(x'')$, where the lift moves the centred representative into the hexagonal cell of the lattice $2^W\mathbb{Z}[\omega]$ with at most one lattice step.
The zero class gets the digit $2^W$, so the core never adds the identity.
An exhaustive check at $W = 1$ to 7 and $W = 9$ found every digit congruent to its residue, the digit of every nonzero residue of minimal norm in its class, a canonical representative for every orbit, and a table of $(4^W-4)/6+2$ entries.
The in-place conversion of $x$ into $(x'',s,j)$ and back costs 340 Toffoli gates and three extra qubits per window.

\paragraph{Digit register.}
The semiclassical QFT corrects the phase of each window by an amount linear in the integer digit, so the circuit computes the digit from $(x'',s,j)$ into two 12-bit registers, applies the rotations and uncomputes the registers.
The digit register costs 532 Toffoli gates at $W = 9$ and 486 at $W = 8$, with a peak of 112 qubits at the start of the window, when about 550 qubits are live.
Its output matched the digit map on every residue at $W = 3$, 4, 5, 6 and 8 and on 16,384 random residues at $W = 9$.

\paragraph{The seven-level register.}
A mixed-radix window over $(3+\omega)2^7$ or $(2-\omega)2^7$ keeps the residue modulo the prime of norm 7 in three qubits that take seven of their eight values.
The circuit prepares the uniform superposition of the seven values with 7 $R_y$ rotations and 8 CNOT gates, as a tree of uniformly controlled rotations~\cite{mottonen2005}.
It measures the register in the Fourier basis of $\mathbb{Z}_7$ with 100 $R_y$ rotations, 184 $R_z$ rotations and 426 CNOT gates, by a Givens elimination of the Fourier transform of order 7 on pairs of basis states that differ in one bit.
Neither circuit uses a Toffoli gate or an ancilla.
State-vector simulations of both window types at 5, 7 and 9 qubits gave each of the $7\cdot4^w$ residues the amplitude $7^{-1/2}2^{-w}$ to within $8\times10^{-17}$, never reached the eighth value, and reproduced the outcome law that the harness samples to within $6\times10^{-15}$.
Negative controls, such as a preparation with one CNOT gate removed, failed these checks.

\paragraph{Read sites of the core.}
The classical addend enters the borrowed entry's gate stream through 4,968 $X$ gates, 3 CNOT gates and 3 $Z$ gates out of 12.4 million operations, at five sites, and no Toffoli gate depends on it.
Our port replaces these gates by select-swap reads of the window's table point at the same five sites.

\paragraph{Counts.}
\cref{tab:counts} adds up each run from its parts.
A unit-orbit window costs 911,737 Toffoli gates at its abort cap plus 532 for its digit register at $W = 9$, and 779,515 plus 486 at $W = 8$.
A mixed-radix window costs 812,589 plus 630 or 632 for its digit register.
The root-sign last addition costs 513,645 at its cap plus 532, 56\% of a unit-orbit window with $W = 9$, which covers 18 exponent bits, and the classical side builds its tables with $310\cdot4^W+1{,}032 = 2^{26.28}$ operations per run at $W = 9$.

\begin{table*}[t]
\centering
\caption{Toffoli gates per run of each configuration at the abort caps, with the digit registers included.}
\label{tab:counts}
\resizebox{\linewidth}{!}{%
\begin{tabular}{lrrr}
\toprule
\textbf{Part} & \textbf{Unit-orbit} & \textbf{Mixed-radix} & \textbf{Mixed-radix and root-sign} \\
\midrule
First lookup & 132,380 (18 bits) & 66,588 (17 bits) & 66,588 (17 bits) \\
Last addition & 912,269 & 912,269 & 514,177 \\
Unit-orbit windows, $W=9$ & $5\times912{,}269$ & 0 & 0 \\
Unit-orbit windows, $W=8$ & $9\times780{,}001$ & $3\times780{,}001$ & $3\times780{,}001$ \\
Mixed-radix window, $(3+\omega)2^7$ & 0 & 813,221 & 813,221 \\
Mixed-radix windows, $(2-\omega)2^7$ & 0 & $10\times813{,}219$ & $10\times813{,}219$ \\
Sine-state preparations & 48 & 45 & 45 \\
\midrule
Run & 12,626,051 & 12,264,316 & 11,866,224 \\
Peak qubits & 1,246 & 1,246 & 1,246 \\
\bottomrule
\end{tabular}}
\end{table*}

\section{End-to-end trials by size}
\label{app:trials}

\cref{tab:trials} lists the pre-registered trials of each configuration at every size.
Every trial succeeded except one root-sign trial at 128 bits, which \cref{sec:eval-size} describes.
Below 256 bits, 1.3 to 8.1\% of the runs were bad or flagged, and at 256 bits 9.2 to 9.9\%.

\begin{table*}[t]
\centering
\caption{End-to-end trials by curve size. Below 256 bits the additions are stand-ins with the failure constants of ordinary windows. Work is the base-2 logarithm of the classical operations of one key recovery.}
\label{tab:trials}
\small
\begin{tabular}{rrrrrr}
\toprule
\textbf{Bits} & \textbf{Solved} & \textbf{Runs} & \textbf{Bad or flagged runs} & \textbf{Work, mean} & \textbf{Work, largest} \\
\midrule
\multicolumn{6}{l}{\emph{Unit-orbit}} \\
16 & 200 of 200 & 16 & 61 of 3,200 (1.9\%) & 18.8 & 20.8 \\
24 & 200 of 200 & 18 & 48 of 3,600 (1.3\%) & 21.9 & 22.2 \\
32 & 200 of 200 & 20 & 119 of 4,000 (3.0\%) & 21.4 & 25.9 \\
64 & 200 of 200 & 23 & 274 of 4,600 (6.0\%) & 26.2 & 31.5 \\
128 & 200 of 200 & 28 & 352 of 5,600 (6.3\%) & 27.5 & 32.9 \\
256 & 563 of 563 & 79 & 4,405 of 44,477 (9.9\%) & 32.7 & 39.06 \\
\midrule
\multicolumn{6}{l}{\emph{Mixed-radix}} \\
16 & 200 of 200 & 16 & 183 of 3,200 (5.7\%) & 21.5 & 25.48 \\
24 & 200 of 200 & 18 & 268 of 3,600 (7.4\%) & 24.2 & 30.42 \\
32 & 200 of 200 & 20 & 225 of 4,000 (5.6\%) & 23.6 & 28.95 \\
64 & 200 of 200 & 23 & 287 of 4,600 (6.2\%) & 24.7 & 29.13 \\
128 & 200 of 200 & 28 & 444 of 5,600 (7.9\%) & 29.3 & 34.40 \\
256 & 563 of 563 & 110 & 6,046 of 61,930 (9.8\%) & 33.3 & 39.85 \\
\midrule
\multicolumn{6}{l}{\emph{Mixed-radix and root-sign}} \\
16 & 200 of 200 & 16 & 194 of 3,200 (6.1\%) & 21.4 & 26.31 \\
24 & 200 of 200 & 18 & 246 of 3,600 (6.8\%) & 23.9 & 29.21 \\
32 & 200 of 200 & 20 & 249 of 4,000 (6.2\%) & 24.0 & 29.24 \\
64 & 200 of 200 & 23 & 296 of 4,600 (6.4\%) & 26.0 & 32.93 \\
128 & 199 of 200 & 28 & 456 of 5,600 (8.1\%) & 33.1 & 39.98 \\
256 & 563 of 563 & 102 & 5,302 of 57,426 (9.2\%) & 34.0 & 37.91 \\
\bottomrule
\end{tabular}
\end{table*}

\section{Directions that do not help}
\label{app:closed}

\cref{tab:closed,tab:closed2} list the directions that we examined, grouped by the part of the run that they change, with the result that closed each one and the grade of its evidence.
A proof is a theorem of \cref{sec:limit} or \cref{app:proofs}, and an exact toy is an exact computation on the small groups and boxes that we tested.
A fitted model is a model fitted to such exact data, a count is a gate count, and a bound limits counted or modeled costs.
An estimate comes from a cost model, measured means circuits or harness trials, and an argument is reasoning that we have not written as a proof.

\begin{table*}[p]
\centering
\caption{Directions in the run's shape, its instances and its measurement that do not shrink the largest run, with the result that closed each one and the grade of its evidence. $B$ is the volume of the largest run.}
\label{tab:closed}
\footnotesize
\begin{tabular}{p{0.40\linewidth}p{0.38\linewidth}p{0.14\linewidth}}
\toprule
\textbf{Direction} & \textbf{Result} & \textbf{Evidence} \\
\midrule
\multicolumn{3}{l}{\emph{Run shape and domain}} \\
Class outputs ($x$-coordinate, unit classes) & at most $\log_2 6 < 2.6$ bits per run (\cref{cor:class}) & proof \\
Four-generator boxes, hexagonal domains & same bound at equal $|S|$, no gain on tested toys & proof, exact toy \\
Adaptive chains of runs & no gain on the tested chains & exact toy, fitted model \\
Smaller later runs in a chain & $T$ 30 to 40\% lower in the ideal model, $B$ unchanged & argument \\
Mid-run group-register measurement, one-register runs & fewer runs at best, no smaller run & argument \\
Short-multiplier trick, postselection on a flag & fewer runs at best, no smaller run & argument \\
\midrule
\multicolumn{3}{l}{\emph{Reductions to related instances}} \\
Several instances per run, self-reductions, twists & \cref{thm:info} applies & proof \\
Multidimensional and multi-twist runs at 256 bits & no better than Eker{\aa}'s tradeoff with our windows & bound \\
Six-unit orbit of the key as the decoder's list & $K$ six times larger, nothing more & bound \\
Endomorphism of norm 7 inside the circuit & 1.3 million Toffoli against 57,000 for a lookup & count \\
\midrule
\multicolumn{3}{l}{\emph{Measurements}} \\
Zero-padding before the QFT & below 0.005 bits on tested boxes, $c = 1$ to 64 & exact toy \\
Measurements other than the QFT & QFT within 2.4 bits of the bound at $c = 64$, 256 & exact toy \\
\bottomrule
\end{tabular}
\end{table*}

\begin{table*}[p]
\centering
\caption{Directions in the arithmetic, the borrowed core and the decoder that do not shrink the largest run, with the result that closed each one and the grade of its evidence. $B$ is the volume of the largest run.}
\label{tab:closed2}
\footnotesize
\begin{tabular}{p{0.40\linewidth}p{0.38\linewidth}p{0.14\linewidth}}
\toprule
\textbf{Direction} & \textbf{Result} & \textbf{Evidence} \\
\midrule
\multicolumn{3}{l}{\emph{Inversion arithmetic}} \\
Our own inversion-based point addition & floor of 0.33 to $0.35\,B_0$ at 18 additions & bound \\
Compressed or recomputed transcripts & at most 9\% compressible, flat qubit-Toffoli product & argument \\
Transcript-free Euclid, Lehmer, half-GCD, divsteps & 1.3 to 4.4 times the cost per pass & count \\
Fermat inversion & at least 10 million Toffoli per division & count \\
Residue-number or unreduced walk values & 514 more qubits, a peak of 1,741 & bound \\
Low-width inversion & 8.7 million Toffoli per inversion & count \\
\midrule
\multicolumn{3}{l}{\emph{Fusion and representation}} \\
Fused divisions, coordinate pairs, shared walks & one gadget saved, 3 to 6 multiplications paid & bound \\
Projective or scale-free accumulators & no saving & bound \\
Per-window offsets on unit-orbit or signed tables & unit moves the offset, unsigned tables 15\% costlier & estimate \\
\midrule
\multicolumn{3}{l}{\emph{Approximate arithmetic and width}} \\
Shorter walk caps, narrower schedules, other adders & at most 4 to 7\% of $B$ & bound \\
Four times the failure budget of the core's walks & 1.5 to 2\% of $B$ & measured \\
Register reuse and held reads & 0.4\% of $B$ & bound, measured \\
\midrule
\multicolumn{3}{l}{\emph{The borrowed core}} \\
Cuts outside the core's walks & at most about 3\% of $B$ & bound, measured \\
Lower failure rates by edits of tables and schedule & no reduction found & measured \\
\midrule
\multicolumn{3}{l}{\emph{Decoder, windows and run count}} \\
Runs of 266 exponent bits & 4 to 8 runs over the budget & measured \\
Runs of 264 exponent bits & not decodable within $2^{40}$ operations & measured \\
A register $b$ of 8 bits & 159 runs, above the budget & measured \\
Mixed-radix windows over other primes & costlier than norm 7 & estimate \\
Shaped windows under the sine top & worse than plain windows & measured \\
Last-addition size, window mix, norm, padding, decoder, flag & 0.3\% of $B$ or less each & estimate, bound \\
\bottomrule
\end{tabular}
\end{table*}

\clearpage
\bibliography{refs}

\end{document}